\documentstyle[11pt,aaspp4,epsfig]{article}  
\lefthead{Moskalenko and Strong}
\righthead{Cosmic-ray positrons and electrons}

\begin{document}

\title{Production and propagation of cosmic-ray positrons and electrons}

\author{I.V.~Moskalenko\altaffilmark{1} and A.W.~Strong}
\affil{Max-Planck-Institut f\"ur Extraterrestrische Physik,
   Postfach 1603, D-85740 Garching, Germany}
\altaffiltext{1}{On leave from Institute for Nuclear Physics, %
  M.V.Lomonosov Moscow State University, 119 899 Moscow, Russia}

\authoremail{imos@mpe-garching.mpg.de; aws@mpe-garching.mpg.de}

\begin{abstract}

We have made a new calculation of the cosmic-ray secondary positron
spectrum using a diffusive halo model for Galactic cosmic-ray
propagation.  The code computes self-consistently the spectra of primary
and secondary nucleons, primary electrons, and secondary positrons and
electrons.  The models are first adjusted to agree with the observed
cosmic-ray Boron/Carbon ratio, and the interstellar proton and Helium
spectra are then computed; these spectra are used to obtain the source
function for the secondary positrons/electrons which are finally
propagated with the same model parameters. The primary electron spectrum
is  evaluated, again  using the  same model.  Fragmentation and energy
losses are computed using realistic distributions for the interstellar
gas and radiation fields, and diffusive reacceleration is also
incorporated.  Our study includes a critical re-evaluation of the
secondary decay calculation for positrons.

The predicted positron fraction is in good agreement with the
measurements up to 10 GeV, beyond which the observed flux is higher than
that calculated.  Since the positron fraction is now accurately measured
in the 1--10 GeV range our primary electron spectrum should be a good
estimate of the true interstellar spectrum in this range, of interest
for gamma ray and solar modulation studies.  We further show that a
harder interstellar nucleon spectrum, similar to that suggested to
explain EGRET diffuse Galactic gamma ray observations above 1 GeV, can
reproduce the  positron observations above 10 GeV without requiring a
primary positron component.

\end{abstract}
\keywords{cosmic rays --- diffusion --- elementary particles --- 
Galaxy: general --- ISM: abundances --- ISM: general}

\section{Introduction}

Secondary positrons in galactic cosmic rays are an important diagnostic
for models of cosmic-ray propagation and also for solar modulation
studies. Recently several new experiments have provided improved data on
both the positron to electron ratio and the positron spectrum itself
(\cite{Barwick97,Barbiellini96,Golden96}). These and previous data have
mainly been compared with model predictions using leaky-box and
diffusion models from Protheroe (1982). Since these predictions were
made,  new information and ideas have contributed to our understanding
of cosmic-ray propagation: new measurements of nucleon
secondary-to-primary ratios, the energy dependence of the propagation,
results from gamma-ray astronomy and studies of the effects of diffusive
reacceleration. For this reason new calculations of secondary positrons
are desirable.

Recently an extensive computer code for the calculation of galactic
cosmic-ray propagation has been developed (\cite{StrongMoskalenko97a}),
which is a further development of the approach described by Strong \&
Youssefi (1995) and Strong (1996). Primary and secondary nucleons,
primary and secondary electrons, and secondary positrons are included.
The basic spatial propagation mechanisms are (momentum-dependent)
diffusion and convection, while in momentum space energy loss and
diffusive reacceleration are treated. Fragmentation and energy losses
are computed using realistic distributions for the interstellar gas and
radiation fields.

The main motivation for developing this code was the prediction of
diffuse Galactic gamma rays for comparison with data from the Gamma Ray
Observatory CGRO instruments EGRET, COMPTEL and OSSE.  More generally
the idea is to develop a model which self-consistently reproduces
observational data of many kinds related to cosmic-ray origin and
propagation: directly via measurements of nuclei, electrons and
positrons, indirectly via gamma rays and synchrotron radiation. These
data provide many independent constraints on any model and our approach
is able to take advantage of this since it must be consistent with all
types of observation.  We emphasize also the use of realistic
astrophysical input (e.g. for the gas distribution) as well as
theoretical developments (e.g. reacceleration).  The code is
sufficiently general that new physical effects can be introduced as
required. In this paper we focus on positrons.  The secondary positron
problem is closely connected with that of gamma-rays from $\pi^0$-decay
since the same cosmic-ray nucleons are involved. For this reason the
present application is a natural extension of the original code.

The basic procedure is first to obtain a set of propagation parameters
which reproduce the cosmic-ray B/C ratio and the spectrum of primary
electrons. The proton and Helium spectra are then computed using these
parameters, and used to obtain the source function for the secondary
positrons and electrons. The  secondary positron and electron spectra
are then computed using the same propagation model. Note that although
the ratio of positrons to electrons is normally used for the comparison
with data, it is in fact the positron spectrum itself which is of
interest in testing propagation models; however since the ratio is more
easily measured experimentally we also use it in this work.

Turning the argument around, we note that if we can compute the positron
spectrum reliably, then accurate measurements of the positron fraction
enable us to determine the true interstellar electron spectrum unless
there is significant charge-sign dependent solar modulation (for a
discussion of the charge-sign dependence see \cite{Clem96}).  This
approach also provides an independent test for models of the primary
electron spectrum, complementing gamma-ray and synchrotron measurements.
Especially around 1 GeV the positron fraction is now rather accurately
known, so the interstellar electron flux at this energy can be
determined.
 
We note that similar calculations of the electron spectrum have been
made by Porter \& Protheroe (1997) using a (1-D) Monte Carlo approach.
They also address the galactic diffuse gamma-ray spectrum. However,
since the main subject of the present paper is positrons, a detailed
comparison of our electron and gamma-ray spectra with theirs will be
made elsewhere (\cite{StrongMoskalenko97b}).

\section{Description of the models}
The models will be described in full detail elsewhere
(\cite{StrongMoskalenko97b}; see also the information in Section 7);
here we summarize briefly their basic features.  The models are three
dimensional with cylindrical symmetry in the Galaxy, and the basic
coordinates are $(R,z,p)$ where $R$ is Galactocentric radius, $z$ is the
distance from the Galactic plane and $p$ is the total particle momentum.
The distance from the Sun to the Galactic centre is taken as 8.5 kpc. In
the models the  propagation region is bounded by $R=R_h$, $z=z_h$ beyond
which free escape is assumed. We take $R_h=30$ kpc. The case $z_h=3$ kpc
has been studied since this is consistent with studies of radioactive
nuclei (the $^{10}$Be/Be ratio: \cite{Lukasiak94}) and synchrotron
radiation.  For a given $z_h$ the diffusion coefficient as a function of
momentum is determined by B/C for the case of no reacceleration; with
reacceleration on the other hand it is the reacceleration strength
(related to the Alfv\'en speed) which is determined by B/C.
Reacceleration provides a natural mechanism to reproduce the B/C ratio
without an ad-hoc form for the diffusion coefficient
(\cite{SimonHeinbach96,HeinbachSimon95,SeoPtuskin94,Letaw93}). The
spatial diffusion coefficient for no reacceleration is taken $\beta D_0$
below rigidity $\rho_0$, $\beta D_0(\rho/\rho_0)^\delta$ above rigidity
$\rho_0$.  The spatial diffusion coefficient with reacceleration assumes
a Kolmogorov spectrum of weak MHD turbulence so $D=\beta
D_0(\rho/\rho_0)^\delta$ with $\delta=1/3$ for all rigidities.  For the
case of reacceleration the momentum-space diffusion coefficient $D_{pp}$
is related to the spatial coefficient using the  formula given by Seo
and Ptuskin (1994) (their equation [9]), and Berezinskii et al.\ (1990).
The main free parameter in this relation is the Alfv\'en speed $v_A$.
The injection spectrum of nucleons is assumed to be a power law in
momentum for  the different species, $dq(p)/dp \propto p^{-\gamma}$ for
the injected {\it density}, corresponding to an injected {\it flux}
$dF(p)/dp \propto \beta p^{-\gamma}$ or $dF(E_k)/dE_k \propto
p^{-\gamma}$ since $dp/dE_k = A/\beta$, where $E_k$ is the kinetic
energy per nucleon, $\beta=v/c$.  The value of $\gamma$ can vary with
species. The interstellar hydrogen distribution uses information from HI
surveys (\cite{GordonBurton76}, \cite{Cox86}) for the atomic component
and CO surveys (\cite{Bronfmann88}) for the molecular hydrogen; we use
also information on the ionized component. The Helium fraction of the
gas is taken as 0.11 by number. The interstellar radiation field for
inverse Compton losses is based on stellar population models and IRAS
(\cite{BloemenDeulThaddeus90}) and COBE (\cite{Boulanger92}) analyses;
in addition we include the cosmic microwave background. The magnetic
field is assumed to have the form $B_\bot=6\,e^{-|z|/5{\rm kpc}-R/20{\rm
kpc}}$ $\mu$G. This magnetic field model is based on recent
observational work (\cite{Vallee94}, \cite{Heiles96}), and our local
field of 4\ $\mu$G is consistent with these papers. The $z$-variation
was chosen to follow studies of radio synchrotron surveys, which suggest
a  vertical extent of emissivity of several kpc (\cite{Phillipps81}) and
hence a comparable extent of both  field and electrons. Energy losses
for electrons by ionization, Coulomb interactions, bremsstrahlung,
inverse Compton and synchrotron are included. The inverse Compton losses
were computed using the Thomson approximation, which is sufficient for
our purposes. The Klein-Nishina corrections affect only electrons of
energies above 100 GeV interacting with optical photon fields, while we
are interested here in electron and positron energies below 100 GeV, and
in any case far-infrared photons dominate the interstellar radiation
energy density. Energy losses for nucleons by ionization and Coulomb
interactions are included following Mannheim \& Schlickeiser (1994). The
distribution of cosmic-ray sources is chosen to reproduce the cosmic-ray
distribution determined by  analysis of EGRET gamma-ray data
(\cite{StrongMattox96}).

First, the primary propagation is computed giving the primary
distribution as a function of ($R, z, p$); then the secondary source
function is obtained from the gas density and cross-sections, and
finally the secondary propagation is computed.  The entire calculation
is performed with momentum as the kinematic variable, since this greatly
facilitates the inclusion of reacceleration. Nucleon spectra are
afterwards converted to $dF(E_k)/dE_k$ for comparison with observations.

\section{The model parameters}
Table~\ref{table1} lists the parameters which were used for the models
without and with reacceleration. The cross-sections for secondary
production from the progenitor $C, N, O$ were taken from Webber, Lee, \&
Gupta (1992), Heinbach \& Simon (1995) and references therein.  The
source relative abundances were taken from Engelmann et al.\ (1990).
The injection spectrum for Carbon was taken as $dq(p)/dp \propto
p^{-2.35}$, for the case of no reacceleration, and $p^{-2.25}$ with
reacceleration.  These values are consistent with Engelmann et al.\
(1990) who give an injection index $2.23\pm0.05$.

\placetable{table1}
\placefigure{fig1}

Figure~\ref{fig1} shows the predicted and observed B/C ratio for the
adopted parameters. We use the Voyager data from Webber et al.\ (1996).
The spectra were modulated to 500 MV appropriate to the Voyager data
using the force-field approximation (\cite{GleesonAxford68}).  The
agreement is sufficiently good for our purpose of computing the
secondary positrons. We can note that below 1 GeV the better fit is
given by the reacceleration model (which also has one less free
parameter), but this has no effect on the positron calculation.
Inclusion of a convective term would be necessary to improve the fit in
this region for the no reacceleration case.

The proton and Helium spectra are computed as a function of $(R,z,p)$ by
the propagation code.  Fig.~\ref{fig2} shows the predicted and observed
proton and Helium spectra. The injection spectrum is adjusted to give a
good fit to the locally measured spectrum, normalizing at 10
GeV/nucleon. For protons we use Seo et al.\ (1991: Fig.~10a) based on
LEAP and IMP8 balloon measurements, and Mori's (1997) `median' flux (his
eq.~[3]). For Helium we use Engelmann et al.\ (1985: Fig.~11) (HEAO-A3),
and the interstellar spectrum given by Seo et al.\ (1991:  Fig.~11a).

\placefigure{fig2}

For the injection spectra of protons, we find $\gamma=2.15$ reproduces
the observed spectra in the case of no reacceleration, and $\gamma=2.25$
with reacceleration.  We find it is necessary to use slightly steeper
(0.2 in the index) injection spectra for  Helium nuclei (see
Table~\ref{table1}) in order to fit the observed spectra in the 1--100
GeV range of interest for positron production. The spectra fit up to
about 100 GeV beyond which the Helium spectrum without reacceleration
becomes too steep and the proton spectrum with reacceleration too flat;
these deviations are of no consequence for the positron calculation.
Although the nucleons are not the main subject of this study, we note
that the reacceleration model reproduces slightly better the observed
spectrum below 10 GeV/nucleon where the interstellar power law in
momentum continues down to 2 GeV/nucleon before bending over
(\cite{Seo91}). The adopted nucleon spectra successfully reproduce the
100--1000 MeV gamma-ray intensity from $\pi^0$-decay
(\cite{Strongetal96,StrongMoskalenko97a,Strong97}), so that we are not
dependent only on the locally measured nucleon spectra for the positron
calculation, but have evidence that it extends thoughout the Galaxy.
Possible gamma-ray evidence for  a flatter nucleon spectrum
(\cite{Hunter97}) and the consequences for positrons are addressed below
(Section 6).

For the primary electrons, an injection index of 2.1 below 10 GeV,
steepening to 2.4 above 10 GeV, was found to reproduce the observed
spectrum up to 30 GeV and is consistent with gamma-ray and synchrotron
radiation studies (\cite{StrongMoskalenko97a}). At higher energies a
further steepening is required but this is not of consequence here. The
flux normalization was chosen to fit direct measurements and the
positron ratio as described in section 5:  $I_e(E)= 3.2\times10^{-8}$
cm$^{-2}$ s$^{-1}$ sr$^{-1}$ MeV$^{-1}$ at 9 GeV.

\section{Production of secondary positrons and electrons}
The production of positrons in collisions of cosmic ray protons with
protons of the interstellar medium has been discussed in detail in
numerous studies (e.g.,
\cite{Stecker70,OrthBuffington76,Protheroe82,Dermer86a}b,
\cite{Murphy87} and reference therein). The muons created through decays
of secondary pions and kaons are fully polarized, which results in
electron/positron decay asymmetry, which in turn causes a difference in
their production spectra.  In the early papers (\cite{OrthBuffington76})
the muon decay asymmetry was treated identically for positrons and
electrons (our paper:  $\xi=-1$ in eq.~[\ref{5.1}]), which leads to
almost a factor 2 lower positron production at high energies. Subsequent
papers appear to have repeated this mistake (e.g., \cite{Protheroe82}).
The difference in $\mu^+$ and $\mu^-$ decay asymmetry as applied to
positron and electron production was noted by Dermer (1986a), but no
attempts to improve the predicted cosmic-ray positron spectrum seem to
have been made. Additionally, the energy spectra of positrons produced
in collisions of isotropic monoenergetic protons with protons at rest
calculated by Murphy, Dermer, \& Ramaty (1987), although in general
agreement with our calculations, still differ in detail. In particular,
they agree better with our calculations using an isotropic distribution
of {\it positrons} in the {\it muon} rest system. Since
positron/electron production and propagation in the Galaxy is among the
hot topics of cosmic-ray physics we give explicitly our formulae for the
positron/electron production spectrum (see Appendix).

\placefigure{fig3}

The energy spectra of positrons ($\xi=+1$ in eq.~[\ref{5.1}]), and
electrons ($\xi=-1$) from the decay of $\pi^\pm$, $K^\pm$ mesons
produced in collisions of isotropic monoenergetic protons with protons
at rest are shown in Fig.~\ref{fig3} for several values of the kinetic
energy of protons. For comparison we show the spectra of positrons
calculated assuming an isotropic distribution in the muon rest system,
$\xi=0$, and also using $\xi=-1$; the latter corresponds to the formula
used by Orth \& Buffington (1976).  Remarkably, the latter one produces
more positrons at maximum (around 30 MeV) as compared to the correct
positron spectrum, while it gives half the  positron yield at high
energies. The curve corresponding to the isotropic distribution lies
exactly in between the two others.

The effect is also clearly seen when integrating over the spectrum of
cosmic-ray protons (Fig.~\ref{fig4}), where we adopted for illustration
the spectrum by Mori (1997), in units of cm$^{-2}$ s$^{-1}$ GeV$^{-1}$:
$J_p = 1.67p_p^{-2.7}(1+[2.5{\rm \ GeV/c}]^2/p_p^2)^{-1/2}$ below 100
GeV, and $J_p = 6.65\times10^{-6}(\varepsilon_p/100{\rm \ GeV})^{-2.75}$
above 100 GeV.  The difference in the positron production spectra for
the $\xi=+1$ and $\xi=-1$ cases reaches a factor 1.6 above $\sim0.2$
GeV. It supports the conclusion made by Orth \& Buffington (1976) that
the neglect of the muon decay asymmetry can result in a 25\% error in
the positron production spectrum, but curiously the inclusion of the
{\it correct} kinematics leads to {\it increase} of the positron yield
at high energies, just the reverse of their inference. The effect of
kaon decay is of minor importance. We included only the most important
channel, $K^\pm \to \mu^\pm + \nu_\mu$, the others would contribute at
the few per cent level. A comparison with the positron production rate
from Protheroe (1982) illustrates the statement above: it generally
agrees with our calculations {\it for the case $\xi=-1$}, also taking
into account that his calculations include a contribution of He nuclei
in cosmic rays and interstellar matter. Our curves are shown for pure
$pp$-interactions, while inclusion of He would give a factor of
$\approx1.4$ increase (e.g., \cite{Dermer86b}). Some discrepancy could
be connected with uncertainties in extrapolation of the inclusive cross
section used by Protheroe at high energies and interaction dynamics. The
uncertainty at low energies (hatched area) is due to uncertainty in the
demodulation of the proton spectrum.

\placefigure{fig4}

Our calculations of the interstellar $e^\pm$ spectra include the
contribution to $\pi^\pm$ production from the channels $p+{\rm He}$,
$\alpha+{\rm H}$, and $\alpha+{\rm He}$. For collisions involving nuclei
with atomic numbers $A>1$, the corresponding cross section is multiplied
by a factor $(A_1^{3/8}+A_2^{3/8}-1)^2$
(\cite{OrthBuffington76,Dermer86a}), while the energy per nucleon is the
kinematic variable.

\section{Interstellar positron and electron spectra}
Fig.~\ref{fig5} shows the computed secondary positron and electron
spectra for the cases without and with reacceleration.  The primary
electron spectrum and the total electron + positron spectrum are  also
shown.  Above a few GeV where solar modulation is small the agreement
with the absolute positron measurements is good within the large
experimental errors. Unfortunately there are few {\it absolute}
measurements of the positron spectrum, and we have to rely mainly on the
positron fraction for comparison.  First, in Fig.~\ref{fig6}, we show
the  results plotted as the positron fraction using  the  total electron
spectrum from Protheroe (1982).  Next, Fig.~\ref{fig7}  shows the
positron fraction but using our {\it computed} interstellar  primary
electron spectrum. None of the computed positron spectra or positron
fractions are seriously in conflict with the observations.  Between 1
and 10 GeV the agreement is very good and the measured data rather
precise.  Above 10 GeV there appears to be an excess above the predicted
ratio although the observational errors are larger. Below a few  GeV
solar modulation will shift the points to lower energy, but  the
implications for modulation models are beyond the scope of this paper.

Reacceleration produces a more peaked positron spectrum with the peak
positron fraction around 600 MeV compared to 400 MeV without
reacceleration.  Although the better fit to the positron fraction is
given by the model without reacceleration (Fig.~\ref{fig6}), this
depends critically on the measured electron spectrum which is not known
to sufficient accuracy to allow a distinction between the models on this
basis. While our non-reacceleration electron spectrum is consistent with
Galactic gamma-rays from 1--1000 MeV and radio synchrotron data down to
38 MHz (\cite{StrongMoskalenko97a,Strong97}),  the reacceleration
spectrum may not be consistent with such data.   It will in any case  be
possible to put critical constraints on electron re-acceleration using
gamma-ray and radio data, and this will be investigated in future work
(\cite{StrongMoskalenko97b}). Note that Protheroe (1982) uses a
radio-based electron spectrum below 1 GeV, but this is increasingly
uncertain at lower energies due to free-free absorption of radio
emission.

\placefigure{fig5}
\placefigure{fig6}
\placefigure{fig7}

Our positron spectrum without reacceleration is steeper than that
computed by Protheroe (1982); ours is a factor 2 higher at 1 GeV and is
about equal to his at 10 GeV. A factor 2 and part of the steeper slope
can be traced to a difference in the production function (see
Fig.~\ref{fig4}), and the remaining difference must be attributed to
many details in the different propagation models (diffusion coefficient,
halo size, gas density, radiation fields, 3-D versus 1-D models). Taking
these reasons into account the calculations are consistent and it is
gratifying that the difference is relatively small. However the new
calculations are able to benefit from better knowledge of the nucleon
spectrum and of cosmic-ray propagation.

Above 10 GeV the predicted positron flux appears too low and the
agreement is not better than for Protheroe (1982).  This would be a hint
for primary positrons {\it or} a harder interstellar nucleon spectrum
(see Sect.~6) than observed locally and emphasizes the importance of
accurate positron measurements in this range.  At energies below 1 GeV
where solar modulation is large our interstellar positron spectrum
should provide a good basis for modulation studies.

As promised in the Introduction, we can use these results to turn the
argument around and  draw conclusions about the {\it interstellar
electron} spectrum using the {\it calculated} secondary positrons and
the {\it measured} positron fraction.  In Fig.~\ref{fig7} we divide the
computed positrons  by the {\it modelled primary electron spectrum} plus
the secondary $e^\pm$;  the agreement with the measured positron
fraction shows that either of the modelled primary electron spectra
(Fig.~\ref{fig5}) are acceptable in the range 0.1--10 GeV.  They are
consistent with the direct measurements of the primary spectrum. The
model without reacceleration is however lower by a factor of about 2
than that required to fit the COMPTEL and EGRET gamma-ray spectrum (see
\cite{StrongMoskalenko97a}); an intermediate spectrum  would
nevertheless be consistent with both positrons and gamma rays within the
uncertainties of each. The model with reacceleration may not be
compatible with gamma-ray and radio data, as mentioned above.

\placefigure{fig8}

\section{A harder interstellar nucleon spectrum ?}
The fit to the positron fraction above 10 GeV could be improved by an
{\it ad hoc} steepening of the electron spectrum, but this would then
disagree with the direct electron measurements which are rather reliable
in this range. Another possibility which we find interesting is to adopt
harder interstellar proton and Helium spectra. Fig.~\ref{fig8}  shows
the proton spectrum and positron fraction for an injection spectral
index $\gamma=2.0$ (model 08--009, no reacceleration).  The primary
electron normalization for this case was adjusted to $I_e(E)=
3.44\times10^{-8}$ cm$^{-2}$ s$^{-1}$ sr$^{-1}$ MeV$^{-1}$ at 9 GeV to
obtain the best overall fit. This model reproduces the positron fraction
well over the whole range; in particular the data above 10 GeV from the
recent HEAT experiment (\cite{Barwick97}) are fitted (note that the
higher positron fractions from earlier experiments can probably be
attributed to the difficulty of distinguishing positrons from protons).
The ambient proton spectral index after propagation is about 2.6 in this
model, compared to the directly measured value of 2.75. This is
especially interesting in view of the independent result from EGRET
diffuse Galactic gamma-ray data (\cite{Hunter97}) that the gamma-ray
spectrum  above 1 GeV is harder than expected for the locally measured
nucleon spectrum (\cite{Mori97,Gralewicz97}). Mori (1997) finds that an
ambient spectral index of 2.41--2.55 is required to fit the gamma-ray
spectrum, which is consistent with that required for the positrons. Thus
our result can be taken as adding some support to the `hard nucleon
spectrum' interpretation of the gamma-ray results, and for the idea that
the Galactic interstellar proton and Helium spectra are harder than
those measured in the heliosphere.  However since this situation is {\it
a priori} unlikely in conventional models of cosmic-ray propagation and
from cosmic-ray anisotropy arguments, it remains an interesting
possibility deserving further investigation and independent tests. If
correct it would explain the positron data without the requirement for a
source of primary positrons.

\section{Conclusions}
We have carried out a new computation of the secondary positron and
primary electron spectra in a self-consistent model of propagation
including nucleons.  The model is more realistic than previous leaky-box
type calculations and incorporates a wider range of astrophysical input.
We have shown that the positron fraction is consistent with measurements
up to 10 GeV, beyond which some excess is apparent.  We have also shown
that it is possible to reverse the normal argumentation to constrain the
interstellar electron spectrum on the basis of the measured positron
fraction and the computed positron spectrum. The resulting interstellar
electron spectrum for 1--10 GeV is consistent with that from direct
measurements. A harder interstellar nucleon spectrum allows the positron
fraction to be fitted also above 10 GeV and would also explain the
high-energy diffuse Galactic gamma-ray spectrum; thus two independent
lines of evidence point to a flatter nucleon spectrum than that directly
measured, so that this possibility has to be taken seriously and further
consequences examined.  At energies below 1 GeV, our calculated positron
spectrum will be of interest for studies of cosmic-ray modulation in the
heliosphere.

More details, including the software and datasets, can be found on \\
{\it http://www.gamma.mpe--garching.mpg.de/$\sim$aws/aws.html} and {\it
$\sim$imos/imos.html}


\section*{APPENDIX}
\appendix

\section{Spectra of secondaries from $pp$-collisions}

The production spectrum of secondary $\gamma$'s, electrons and
positrons can be obtained if one knows the distributions of pions
$F_\pi(\varepsilon_\pi, \varepsilon_p)$ and kaons $F_K(\varepsilon_K,
\varepsilon_p)$ from a collision of a proton of energy $\varepsilon_p$,
and the distribution of secondaries $F_s(\varepsilon_s,
\varepsilon_{\pi,K})$ from the decay of a pion/kaon of energy
$\varepsilon_{\pi,K}$
\begin{equation}
\label{3.1}
j_s(\varepsilon_s) = n_H  \sum_{i=\pi,K}
  \int_{\varepsilon_p^{\min}}^\infty d\varepsilon_p \,
     J_p(\varepsilon_p) \left\langle \eta\sigma_i(\varepsilon_p)\right\rangle
     \int_{\varepsilon_i^{\min}(\varepsilon_s)}
        ^{\varepsilon_i^{\max}(\varepsilon_p)} d\varepsilon_i \,
        F_s(\varepsilon_s,\varepsilon_i)F_i(\varepsilon_i,\varepsilon_p)\, ,
\end{equation}
where $n_H$ is the atomic hydrogen number density, $J_p(\varepsilon_p)$
is the proton flux, and $\left\langle \eta \sigma_{\pi,K}(\varepsilon_p)
\right\rangle$ is the inclusive cross section of pion/kaon production (a
convenient parametrization for different channels is given by
\cite{Badhwar77}, and \cite{Dermer86a}).
The minimum proton energy $\varepsilon_p^{\min}$ that contributes to the
production of a meson with energy $\varepsilon_i^{\min}$, and the
maximal energy of the produced meson
$\varepsilon_i^{\max}(\varepsilon_p)$ can be easily derived from
kinematical considerations, e.g.\ by equating $\sqrt{s}= [2m_p
(\varepsilon_p +m_p)]^{1/2} =\sqrt{m_X^2 +\varepsilon_i^{*2}-m_i^2}
+\varepsilon_i^*$, where $s$ is the square of the total energy in the
center-of-mass system (CMS), $m_X$ depends on the reaction channel, and
$\varepsilon_i^*$ is the CMS energy of the produced meson (connected
with the laboratory system energy $\varepsilon_i$ via the Lorentz
transformation).
The minimum meson energy that contributes to the production
of a secondary with energy $\varepsilon_s$ is given by
$\varepsilon_{\pi^0}^{\min} (\varepsilon_\gamma) =\varepsilon_\gamma
+m_{\pi^0} /(4\varepsilon_\gamma)$ for $\gamma$-rays;
$\varepsilon_{\pi^\pm,K^\pm}^{\min} =m_{\pi^\pm,K^\pm}$ for
$\varepsilon_e \le E\equiv \frac1{2}m_\mu\gamma'_\mu (1 +\beta'_\mu)$, and
$\varepsilon_{\pi^\pm,K^\pm}^{\min}(\varepsilon_e) =\frac1{2}m_{\pi,K}
(\varepsilon_e/E +E/\varepsilon_e)$ if $\varepsilon_e> E$, where
$\gamma'_\mu$, $\beta'_\mu$ are the muon Lorentz factor and speed in
the meson rest system (see below).

\section{Pion production in $pp$-collisions}

We consider pion production in $pp$-collisions following a method
developed by Dermer (1986ab), which combines isobaric (\cite{Stecker70})
and scaling (\cite{Badhwar77,StephensBadhwar81})
models of the reaction. The isobaric model was shown to work well at
low energies, while at high energies the relevant model is based
on scaling arguments. In the transition region we join the models with
a linear connection in the regime between 3 and 7 GeV.

\subsection{Stecker's model}

Assuming that the outgoing $\Delta$-isobar of mass $m_\Delta$ travels
along the initial direction of the colliding protons in the CMS and
decays isotropically, the distribution of the $\pi$'s in the laboratory
system (LS) is
\begin{eqnarray}
\label{4.1}
f_\pi(\varepsilon_\pi, \varepsilon_p; m_\Delta)=
   \frac1{4 m_\pi \gamma'_\pi \beta'_\pi} 
         &&\, \left\{
      \frac1{\gamma^+_\Delta \beta^+_\Delta}
      H[ \gamma_\pi; 
         \gamma^+_\Delta \gamma'_\pi (1 -\beta^+_\Delta \beta'_\pi),
         \gamma^+_\Delta \gamma'_\pi (1 +\beta^+_\Delta \beta'_\pi) ]
\right.
\nonumber \\
&&
\left.
        +\frac1{\gamma^-_\Delta \beta^-_\Delta}
      H[ \gamma_\pi; 
         \gamma^-_\Delta \gamma'_\pi (1 -\beta^-_\Delta \beta'_\pi),
         \gamma^-_\Delta \gamma'_\pi (1 +\beta^-_\Delta \beta'_\pi) ]
   \right\},
\end{eqnarray}
where $H[x;a,b]=1$ if $a\le x\le b$ and = 0 otherwise, $\varepsilon_\pi$
is the pion energy, and $\varepsilon_p$ is the LS energy of the colliding
proton. The Lorentz factors of the forward (+) and backward (--) moving
isobars in the LS are $\gamma^\pm_\Delta =\gamma_c \gamma^*_\Delta
(1 \pm\beta_c \beta^*_\Delta)$, where $\gamma_c=\sqrt{s}/2m_p$ is the 
Lorentz factor of the CMS in the LS, $\gamma^*_\Delta =(s +m^2_\Delta 
-m_p) /2\sqrt{s}m_\Delta$ is the Lorentz factor of the isobar in the CMS.
The pion Lorentz factor in the rest frame of the $\Delta$-isobar is
$\gamma'_\pi =(m^2_\Delta +m^2_\pi -m^2_p) /2m_\Delta m_\pi$.

Integration over the isobar mass spectrum (Breit-Wigner distribution)
yields the distribution of pions
\begin{eqnarray}
\label{4.2}
F_\pi(\varepsilon_\pi, \varepsilon_p)=
   \Gamma \left\{
      \tan^{-1}\left(\frac {\sqrt{s} -m_p -m^0_\Delta} \Gamma \right)
\right. && \left. 
     -\tan^{-1}\left(\frac {m_p +m_\pi -m^0_\Delta} \Gamma \right)
   \right\}^{-1}
\nonumber \\
&& \times
   \int_{m_p+m_\pi}^{\sqrt{s}-m_p} dm_\Delta \,
      \frac {f_\pi(\varepsilon_\pi, \varepsilon_p; m_\Delta)}
         {(m_\Delta -m^0_\Delta)^2 +\Gamma^2}\, ,
\end{eqnarray}
where $m^0_\Delta$ is the average mass of the $\Delta$-isobar, and
$\Gamma$ is the width of the Breit-Wigner distribution (note that
$\Gamma$ here is a factor of 2 higher than the value usually given in
particle data tables).

\subsection{The scaling model}

The Lorentz invariant cross sections for charged and neutral pion
production in $pp$-collisions inferred from experimental data at
$\varepsilon_p\ga13.5$ GeV are given by Badhwar, Stephens, \& Golden
(1977), and Stephens \& Badhwar (1981)
\begin{equation}
\label{4.3}
\varepsilon_\pi \frac{d^3\sigma}{d^3 p_\pi}=
A G_\pi(\varepsilon_p) (1-\tilde{x}_\pi)^Q \exp[-Bp_\perp/(1+4m_p^2/s)]\, ,
\end{equation}
where
\begin{eqnarray}
\label{4.4}
&& G_{\pi^\pm}(\varepsilon_p) = (1 +4m_p^2/s)^{-R}, \\
&& G_{\pi^0}  (\varepsilon_p) = (1 +23\varepsilon_p^{-2.6})(1 -4m_p^2/s)^R, \\
&& Q            = (C_1 -C_2 p_\perp +C_3 p_\perp^2)/\sqrt{1+4m_p^2/s}\, , \\
&& \tilde{x}_\pi= \sqrt{x_\|^* +(4/s)(p_\perp^2 +m_\pi^2)}\, , \\
&& x_\|^*       = \frac 
   {2 m_\pi \sqrt{s}\, \gamma_c \gamma_\pi (\beta_\pi \cos \theta -\beta_c)}
   {[(s -m_\pi^2 -m_X^2)^2 -4 m_\pi^2 m_X^2]^{1/2}}\, ,
\end{eqnarray}
$\theta$ is the pion LS polar angle, 
$A$, $B$, $C_{1,2,3}$, $R$ are
the positive constants, and $m_X$ depends on the reaction channel:
$m_X=2m_p$ for reaction $pp\to\pi^0X$ , $m_X=m_p+m_n$ for
$pp\to\pi^+X$, $m_X=m_d$ for $pp\to\pi^+d$, $m_X=2m_p+m_\pi$ for
$pp\to\pi^-X$, $m_X=m_p+m_n$ for $pp\to K^+X$, and $m_X=2m_p+m_K$ for
$pp\to K^-X$.

The LS energy distribution of pions can be obtained by integration over
the LS polar angle
\begin{equation}
\label{4.5}
F_\pi(\varepsilon_\pi, \varepsilon_p)=
   \frac{2 \pi p_\pi}
   {\left\langle \eta\sigma_\pi(\varepsilon_p) \right\rangle_{sm}}
   \int^1_{\cos \theta_{\min} } d\cos \theta 
      \left(\varepsilon_\pi \frac{d^3\sigma}{d^3 p_\pi}\right)\, ,
\end{equation}
where, provided $-1\le \cos \theta_{\min} \le1$, 
\begin{equation}
\label{4.6}
\cos \theta_{\min} =\frac1 {\beta_c \gamma_c p_\pi}
   \left(\gamma_c \varepsilon_\pi -\frac{s -m_X^2 +m_\pi^2}{2\sqrt{s}}
   \right)\, ,
\end{equation}
and $\left\langle \eta \sigma_\pi(\varepsilon_p) \right\rangle_{sm}$ is
the inclusive cross section of pion production in the scaling model.

The charged kaon production is of minor importance in comparison
with pion production. The parameters of the Lorentz invariant cross section
\begin{equation}
\label{4.7}
\varepsilon_K \frac{d^3\sigma}{d^3 p_K}= A (1-\tilde{x}_K)^C \exp(-Bp_\perp)
\end{equation}
are given by Badhwar, Stephens, \&
Golden (1977).  The
two main decay modes are $K^\pm \to \mu \nu_\mu$ (63.5\%) and $K^\pm
\to \pi^0 \pi^\pm$ (21.2\%), other modes, with three particles in the
final state, contribute at a few per cent level (\cite{ParticleData90}).

\section{Pion decay}

The muons from the decay of charged pions are created fully polarized,
which results in $e^\pm$ decay asymmetry. The secondary
electron/positron distribution in the muon rest system assuming a
massless electron/positron is given by (\cite{ParticleData90})
\begin{equation}
\label{5.1}
f'_\xi(\varepsilon'_e,\cos \theta')
   =\frac{8{\varepsilon'}_e^2}{m_\mu^3}
      \left[3 -4\frac{\varepsilon'}{m_\mu}
      -\xi\cos \theta'\left(1 -4\frac{\varepsilon'}{m_\mu}\right)\right],
\end{equation}
where $\xi=\pm1$ for $e^\pm$, $\xi=0$ for the isotropic distribution,
$\varepsilon'_e$, $\theta'$ are the electron/positron energy in the
muon rest system and polar angle respectively.


In the LS, the electron/positron distribution is
\begin{equation}
\label{5.2}
f_\xi(\varepsilon_e,\varepsilon_\mu)=\int_{\cos \theta_L}^1 d\cos \theta \,
   f'_\xi(\varepsilon'_e,\cos \theta')
   J\left(\frac{\varepsilon'_e,\cos \theta'}
          {\varepsilon_e,\cos \theta}\right),
\end{equation}
where $\cos \theta_L=\max\{-1,(1-m_\mu/2\varepsilon _e\gamma_\mu )/
\beta_\mu \}$, $\gamma_\mu$, $\beta_\mu$ are the muon Lorentz factor
and speed in LS, $J$ is the Jacobian
\begin{equation}
\label{5.3}
J\left(\frac{\varepsilon'_e,\cos \theta'}{\varepsilon_e,\cos \theta}\right)
   =\frac{\varepsilon_e}{\varepsilon'_e}
   =\frac 1{\gamma_\mu (1-\beta_\mu \cos \theta )},
\end{equation}
and
\begin{eqnarray}
\label{5.4}
&& \varepsilon'_e
   =\varepsilon_e\gamma_\mu (1-\beta_\mu \cos \theta),\nonumber\\
&& \cos \theta' 
   =\frac{\cos \theta -\beta_\mu }{1-\beta_\mu \cos \theta }.
\end{eqnarray}

The LS distribution of $\mu^\pm$ from the $\pi^\pm$-decay is
\begin{equation}
\label{5.5}
f_\mu (\varepsilon_\mu,\varepsilon_\pi)=
   \frac1{2m_\mu\gamma_\pi\beta_\pi\gamma'_\mu\beta'_\mu},\
      m_\mu \gamma^-_\mu\leq \varepsilon_\mu \leq m_\mu \gamma_\mu^+,
\end{equation}
where $\varepsilon_\mu$ is the muon LS energy, $\gamma_\mu^\pm=
\gamma_\pi \gamma'_\mu(1\pm \beta_\pi \beta'_\mu)$, $\gamma'_\mu$,
$\beta'_\mu\approx0.2714$ are the muon Lorentz factor and speed in the
pion rest system, and $\gamma_\pi$, $\beta_\pi$ are the pion Lorentz
factor and speed in the LS. Therefore, the LS distribution of $e^\pm$
from the decay of pion with Lorentz factor $\gamma_\pi$ is given by
\pagebreak[1]
\begin{eqnarray}
\label{5.6} 
F_\xi
&&\, 
(\varepsilon_e,\varepsilon_\pi)
    =\frac1{m_\mu \gamma_\pi \beta_\pi \gamma'_\mu \beta'_\mu}
\nonumber \\ 
\times
&&
\left\{
   \begin{array}{ll}
      X_\xi(\gamma^+_\mu)-X_\xi(\gamma^-_\mu),
         &\, 0\leq \varepsilon_e\leq m_\mu /2\gamma^+_\mu(1+\beta^+_\mu) \\
      X_\xi(\gamma_1)-X_\xi(\gamma^-_\mu)
      +Y_\xi(\gamma^+_\mu)-Y_\xi(\gamma_1),
         &\, m_\mu /2\gamma^+_\mu(1+\beta^+_\mu)\leq \varepsilon_e
            \leq m_\mu /2\gamma^-_\mu(1+\beta^-_\mu) \\
      Y_\xi(\gamma^+_\mu)-Y_\xi(\gamma^-_\mu),
         &\, m_\mu /2\gamma^-_\mu(1+\beta^-_\mu)\leq \varepsilon_e
            \leq m_\mu /2\gamma^-_\mu(1-\beta^-_\mu) \\
      Y_\xi(\gamma^+_\mu)-Y_\xi(\gamma_1),
         &\, m_\mu /2\gamma^-_\mu(1-\beta^-_\mu)\leq \varepsilon_e
            \leq m_\mu /2\gamma^+_\mu(1-\beta^+_\mu)
\end{array}\right.
\end{eqnarray}
where $\gamma_1 =\frac{\varepsilon_e} {m_\mu } +\frac{m_\mu }
{4\varepsilon_e}$, and
\begin{eqnarray}
\label{5.7}
X_\xi(\gamma) &=& \frac{m_\mu}{2}
   \int^\gamma d\gamma_\mu 
   \int_{-1}^1 d\cos \theta \,
      f'_\xi(\varepsilon_e',\cos \theta')
      J\left(\frac{\varepsilon_e',\cos \theta'}
                   {\varepsilon_e,\cos \theta} \right), \nonumber \\
Y_\xi(\gamma) &=& \frac{m_\mu}{2}
   \int^\gamma d\gamma_\mu 
   \int_{\frac1{\beta_\mu}(1-\frac{m_\mu}{2\varepsilon_e\gamma_\mu})} ^1 
      d\cos\theta \,
      f'_\xi(\varepsilon_e',\cos \theta')
      J\left(\frac{\varepsilon_e',\cos \theta'}
                   {\varepsilon_e,\cos \theta} \right).
\end{eqnarray}
After the integration, one can obtain
\begin{equation}
\label{5.8}
X_\pm(\gamma )=\frac4{9}
   \left(\frac{\varepsilon_e}{m_\mu}\right)^2 \left\{
      \frac{\varepsilon_e}{m_\mu }
         [-32\gamma^3(1\pm \beta ) +\gamma (24\pm 32\beta )] 
      +\gamma^2(27\pm 9\beta )  \mp 9\ln [\gamma (1+\beta )]
   \right\},
\end{equation}
\begin{equation}
\label{5.9}
X_0(\gamma )=
\left(\frac{\varepsilon_e}{m_\mu}\right)^2 \left\{
   \left(\frac{\varepsilon_e}{m_\mu}\right)
      \left(-\frac{128}{9}\gamma^3 +\frac{32}{3}\gamma\right)
   +12\gamma^2
\right\},
\end{equation}
\begin{eqnarray}
\label{5.10}
Y_+(\gamma)
&&\,
=\frac1{12}
\left\{
   \left(\frac{\varepsilon_e} {m_\mu}\right)^3 \left[
      16\ln \left(\frac{\gamma +1}{\gamma -1}\right) -64\gamma(1-\beta)
   \right]
\right.
\nonumber\\
&&
\left.
   +\left(\frac{\varepsilon_e}{m_\mu }\right)^2 \left[
      48\gamma^2(1-\beta) +24\ln \left(\frac\beta{1+\beta }\right)
   \right]
   -2\ln (\gamma\beta) +10\ln[\gamma (1+\beta )]
\right\},
\end{eqnarray}
\begin{eqnarray}
\label{5.11}
Y_-(\gamma)
&&\,
=\frac1{36}
\left\{
   \left(\frac{\varepsilon_e}{m_\mu } \right)^3 \left[
      -512\gamma^3(1-\beta ) +\gamma(576-320\beta ) 
      -48\ln\left(\frac{\gamma +1}{\gamma-1}\right)
   \right]
\right.
\nonumber\\
&&
\left.
   +\left(\frac{\varepsilon_e}{m_\mu }\right)^2 \left[
      288\gamma^2(1-\beta) -72\ln \left(\frac \beta {1+\beta }\right)
   \right]
  +6\ln (\gamma\beta )+30\ln [\gamma(1+\beta ) ]
\right\}.
\end{eqnarray}
\begin{eqnarray}
\label{5.12}
Y_0(\gamma)
=\frac1{18}
\left\{ 
   \left(\frac{\varepsilon_e}{m_\mu} \right)^3
      [-128\gamma^3(1-\beta) 
\right.
&&\,
      +\gamma(96-32\beta)]
\nonumber\\
&&
\left.
   +\left(\frac{\varepsilon_e} {m_\mu}\right)^2
      108\gamma^2(1-\beta)
   +15\ln[\gamma(1+\beta)]
\right\},
\end{eqnarray}
At large Lorentz factors, $\gamma\ga20$, it is necessary to use the
series expansions.

The distribution of $\gamma$-rays from $\pi^0$ decay is given by
\begin{equation}
\label{5.0}
F_\gamma(\varepsilon_\gamma,\varepsilon_{\pi^0})=\frac{2}{p_\pi}, \
   \frac1{2} m_\pi\gamma_\pi (1 -\beta_\pi) \leq \varepsilon_\gamma \leq
   \frac1{2} m_\pi\gamma_\pi (1 +\beta_\pi),
\end{equation}
where the factor two accounts for two photons per  decay.

\section{Kaon decay}

The two main kaon decay modes are $K^\pm \to \mu \nu_\mu$ (63.5\%) and
$K^\pm \to \pi^0 \pi^\pm$ (21.2\%), and thus they also contribute to
the secondary $e^\pm$ spectrum. The first mode is similar to $\pi^\pm$
decay, and all formulas presented in previous section are valid if one
replaces index $\pi$ with $K$ and takes appropriate values for
$\gamma'_\mu$, $\beta'_\mu$.

The charged pion distribution from the two-pion decay of the kaon is
\begin{equation}
\label{6.1}
f_\pi (\varepsilon_\pi,\varepsilon_K)=
   \frac1{2m_\pi\gamma_K \beta_K \gamma'_\pi \beta'_\pi},\
      m_\pi \gamma^-_\pi\leq \varepsilon_\pi \leq m_\pi \gamma_\pi^+,
\end{equation}
where $\varepsilon_\pi$ is the pion LS energy, $\gamma_\pi^\pm=
\gamma_K \gamma'_\pi(1\pm \beta_K \beta'_\pi)$, $\gamma'_{\pi^\pm}$,
$\beta'_{\pi^\pm}\approx0.8267$ are the charged pion Lorentz factor and
speed in the kaon rest system, and $\gamma_K$, $\beta_K$ are the kaon
Lorentz factor and speed in the LS. Therefore the LS distribution of
pions from the reaction $pp\to K^\pm X$ is given by
\begin{equation}
\label{6.2}
F_\pi(\varepsilon_\pi, \varepsilon_p)=
   \frac{\pi m_K}{m_\pi \gamma'_\pi \beta'_\pi}
      \left\langle \eta \sigma_K(\varepsilon_p) \right\rangle^{-1}
   \int^\infty_{\varepsilon_K^{\min}(\varepsilon_\pi)} d\varepsilon_K
   \int^1_{\cos \theta_{\min} } d\cos \theta 
      \left(\varepsilon_K \frac{d^3\sigma}{d^3 p_K}\right)\, ,
\end{equation}
where $\cos \theta_{\min}$ is given by eq.~(\ref{4.6}) with the
replacement of the index $\pi$ with $K$, and
$\varepsilon_K^{\min}(\varepsilon_\pi)$ can be obtained from the
equation $\varepsilon_\pi=\gamma'_\pi(\varepsilon_K^{\min}+ \beta'_\pi
p_K^{\min})$.

\clearpage
 
\begin{deluxetable}{cccccccccccc}
\tablecolumns{12}
\footnotesize
\tablecaption{Parameters of models.  \label{table1}}
\tablehead{
\colhead{} & \colhead{} & \colhead{} & 
\colhead{} & \colhead{} & \colhead{} & 
\multicolumn{3}{c}{protons} & \multicolumn{3}{c}{Helium} \\
\cline{7-9} \cline{10-12} \\
\colhead{Model} & \colhead{$z_h$, kpc} & \colhead{$D_0$, cm${^2}$/s} &
\colhead{$\rho_0$, MV/c} & \colhead{$\delta$} & \colhead{$v_A$, km/s} &
\colhead{$\gamma$} & \colhead{$p_0$\tablenotemark{a}} &
\colhead{$I_0$\tablenotemark{b}} &
\colhead{$\gamma$} & \colhead{$p_0$\tablenotemark{a}} &
\colhead{$I_0$\tablenotemark{b}} }
\startdata

08--005 & 3  & $2.0\times10^{28}$ & $3.0\times10^3$ & 0.60 & 0  & 2.15 
& 10$^4$ & $3\times10^{-6}$ & 2.35 & $4\times10^4$ & $4\times10^{-8}$ \nl

08--006 & 3  & $4.2\times10^{28}$ & $3.0\times10^3$ & 0.33 & 20 & 2.25 
& 10$^4$ & $3\times10^{-6}$ & 2.45 & $4\times10^4$ & $4\times10^{-8}$ \nl

08--009 & 3   & $2.0\times10^{28}$ & $3.0\times10^3$ & 0.60 & 0  & 2.00
& 10$^4$ & $3\times10^{-6}$ & 2.00 & $4\times10^4$ & $4\times10^{-8}$ \nl
\enddata

\tablenotetext{a}{ in units MeV c$^{-1}$ nucleus$^{-1}$}
\tablenotetext{b}{ in units cm$^{-2}$ s$^{-1}$ sr$^{-1}$
  (MeV/c/nucleus)$^{-1}$}
\tablecomments{Model 08--005 and 08--009 are without reacceleration,
model 08--006 is with reacceleration. Spatial diffusion coefficient for
no reacceleration:  $\beta D_0$ below rigidity $\rho_0$, $\beta
D_0(\rho/\rho_0)^\delta$ above rigidity $\rho_0$.  Spatial diffusion
coefficient with reacceleration:  $\beta D_0(\rho/\rho_0)^\delta$ for
all rigidities. Nucleon injection density spectrum:  $dq(p)/dp \propto
p^{-\gamma}$. Nucleon spectra normalized to flux $I_0$ at momentum
$p_0$.}

\end{deluxetable}

\clearpage
\def\figwidth{80mm}
\def\fwidth{120mm}
\setlength{\unitlength}{1mm}

\begin{figure}[thb]
   \begin{picture}(165,120)(0,0)
      \put(0,0){\makebox(165,0)[b]%
{\psfig{file=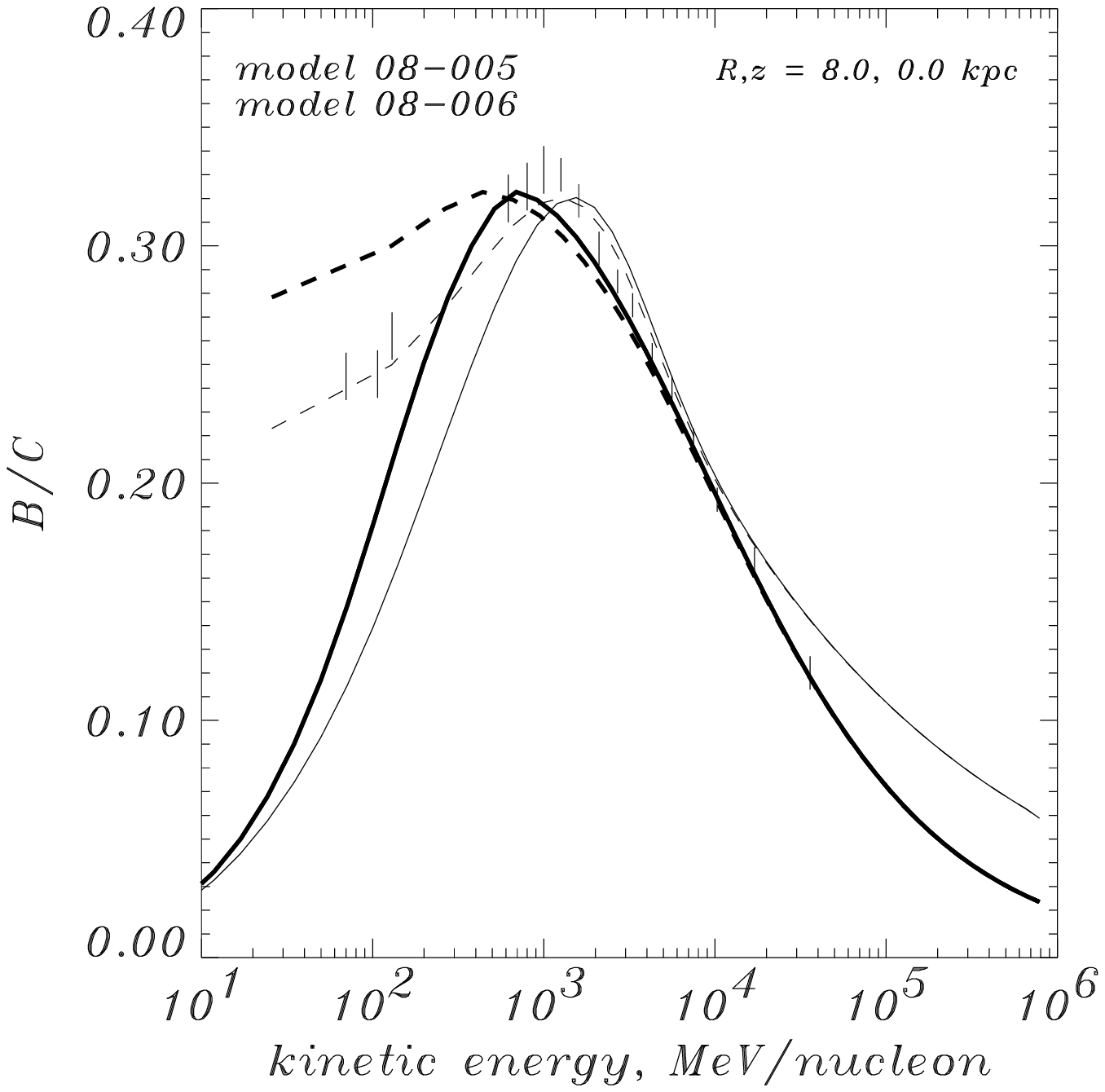,width=\fwidth,clip=}}}
   \end{picture}
\figcaption[fig1.eps]{
B/C ratio. Thick line: model with no reacceleration, nucleon
injection spectrum index 2.35. Thin line: model with  reacceleration,
nucleon injection spectrum index 2.25. Dashed lines: modulated to 500
MV. Data: Voyager, Webber et al.\ (1996).  \label{fig1}}
\end{figure} 

\begin{figure}[thb]
   \begin{picture}(165,80)(0,0)
      \put(0,0){\makebox(85,0)[lb]%
{\psfig{file=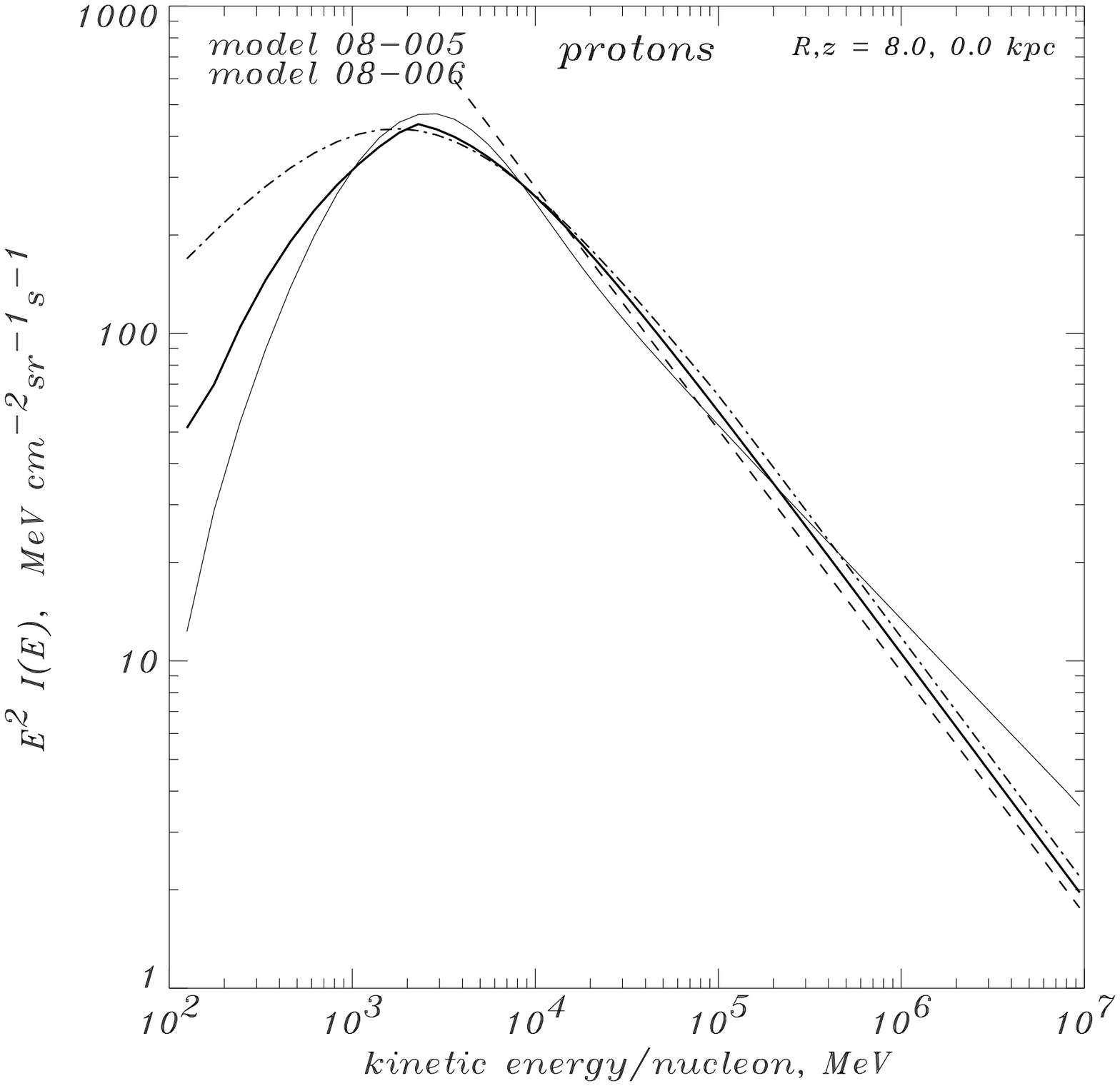,width=\figwidth,clip=}}}
      \put(85,0){\makebox(85,0)[lb]%
{\psfig{file=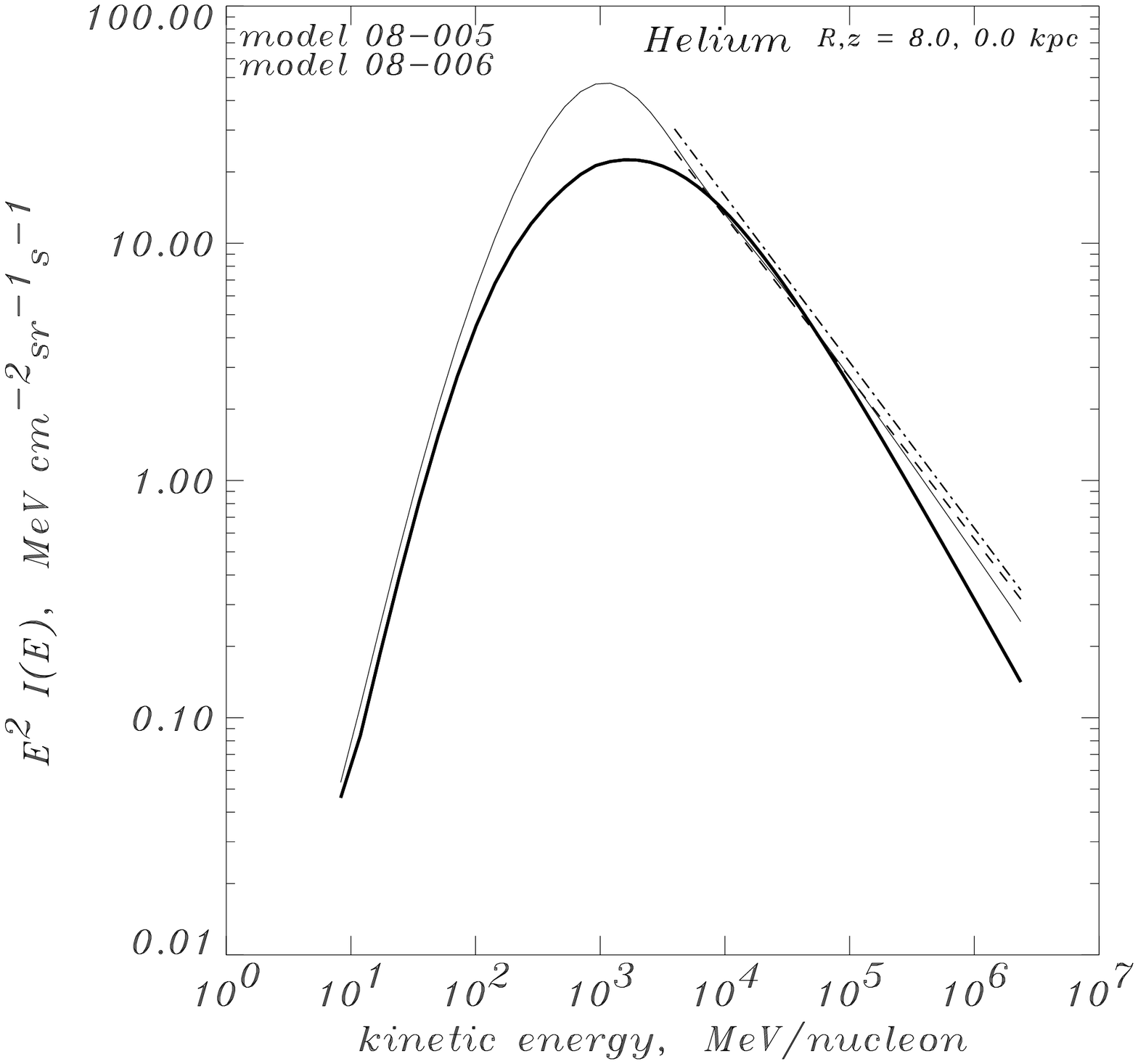,width=\figwidth,clip=}}}
   \end{picture}
\figcaption[fig2a.eps,fig2b.eps]{ 
{\it Left panel}:  spectra of protons. Thick line: model with no
reacceleration, injection spectrum index 2.15. Thin line: model with
reacceleration, injection spectrum index 2.25. Dashed line: Seo et al.\
(1991) `interstellar'; Dash-dotted line:  Mori (1997) median spectrum.
{\it Right panel}: spectra of Helium.  Thick line: model with no
reacceleration, injection spectrum index 2.35. Thin line: model with
reacceleration, injection spectrum index 2.45.  Dashed line: Seo et al.\
(1991) `interstellar'; Dash-dotted line: Engelmann et al.\ (1985).
\label{fig2}}
\end{figure} 

\begin{figure}[thb]
   \begin{picture}(165,80)(0,0)
      \put(0,0){\makebox(85,0)[lb]%
{\psfig{file=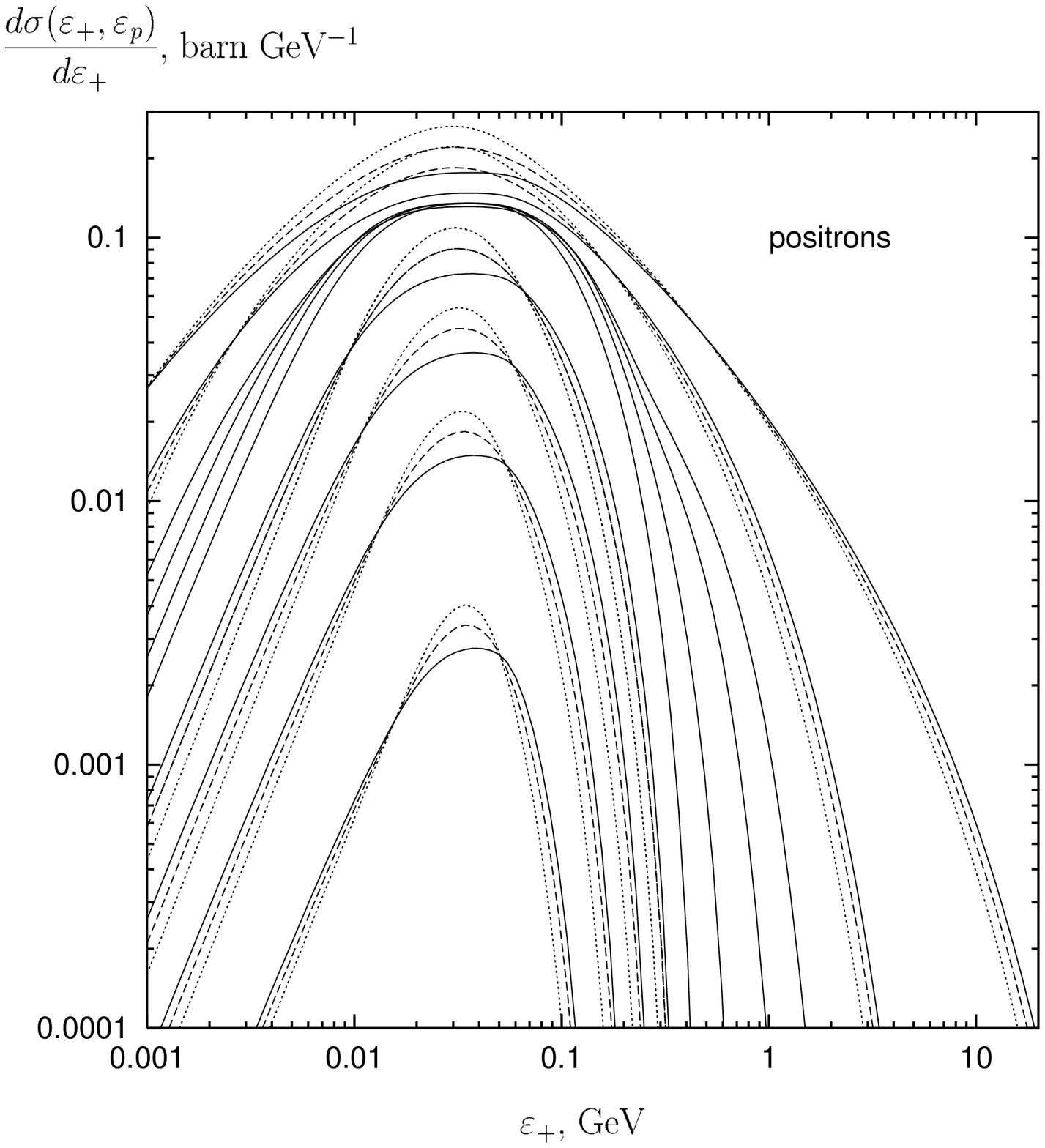,width=\figwidth,clip=}}}
      \put(85,0){\makebox(85,0)[lb]%
{\psfig{file=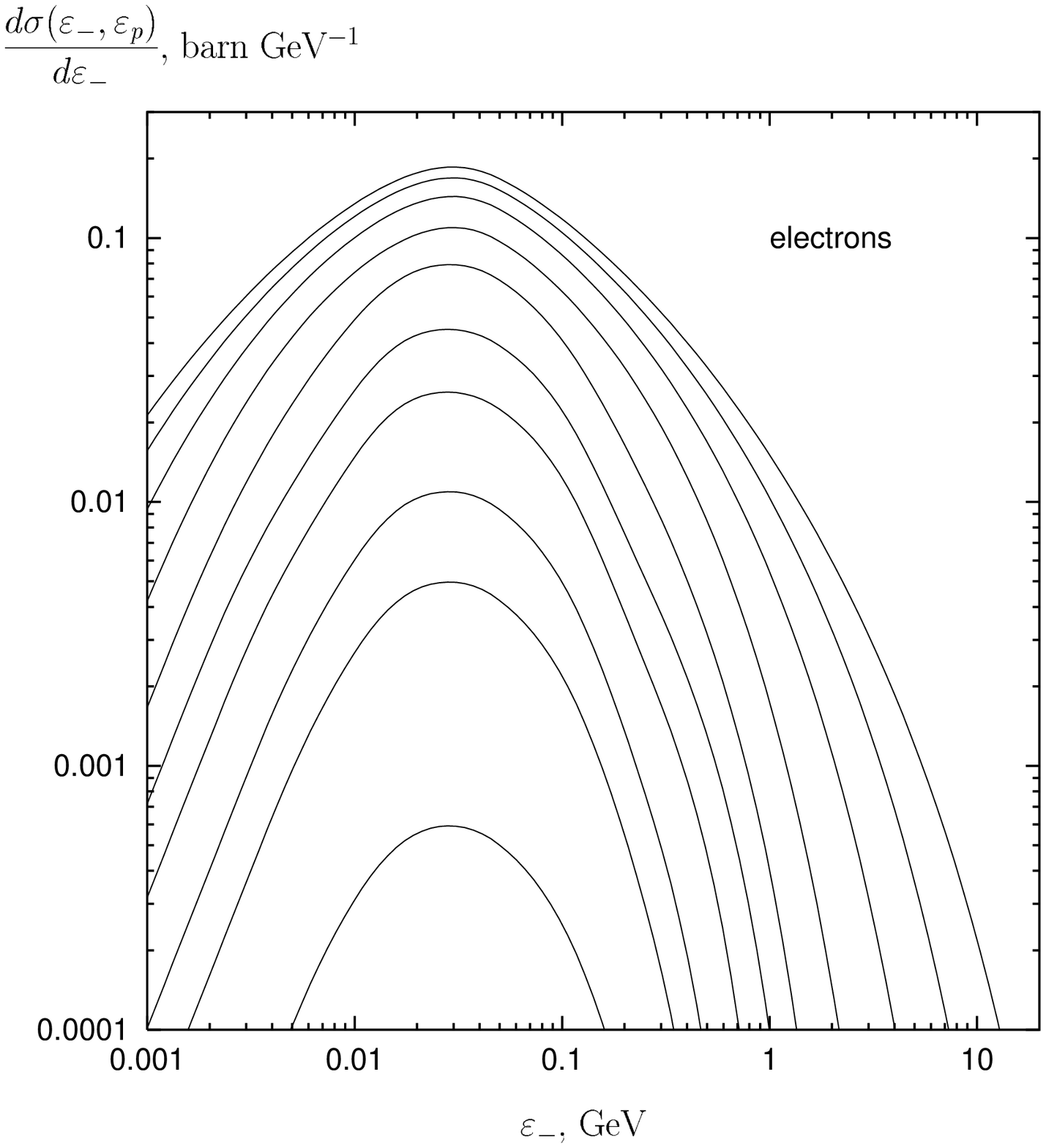,width=\figwidth,clip=}}}
   \end{picture}
\figcaption[fig3a.eps,fig3b.eps]{ 
{\it Left panel}:  solid lines show the energy spectra of positrons from the
decay of $\pi^+$, $K^+$  produced in collisions of isotropic
monoenergetic protons with protons at rest for various proton kinetic
energies (from bottom to top): $\varepsilon_p-m_p=0.316$, 0.383, 0.464,
0.562, 0.681, 1.0, 1.78, 3.16, 10.0, 100.0 GeV ($\xi=+1$,
eq.~[\ref{5.1}]).  Dashed lines show the spectra calculated assuming an
isotropic distribution of $e^+$ in the muon rest system ($\xi=0$) for
some proton kinetic energies  only.  Dotted curves show the spectra
calculated for $\xi=-1$, corresponding to the distribution used by Orth
\& Buffington 1976, their eq.\ [D9].
{\it Right panel}: energy spectra of electrons from the decay of $\pi^-$,
$K^-$  produced in collisions of isotropic monoenergetic protons with
protons at rest for several proton kinetic energies  (from bottom to
top):  1.0, 1.21, 1.47, 2.15, 3.16, 4.64, 10.0, 21.5, 46.4, 100.0 GeV
($\xi=-1$).  \label{fig3}}
\end{figure} 

\begin{figure}[thb]
   \begin{picture}(165,80)(0,0)
      \put(0,0){\makebox(165,0)[b]%
{\psfig{file=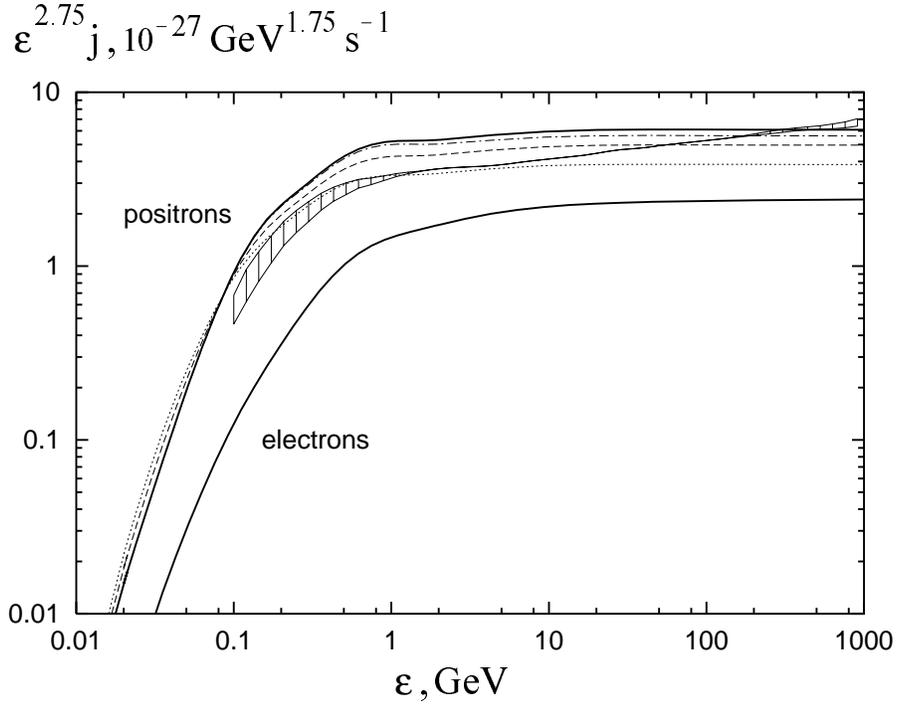,width=\fwidth,clip=}}}
   \end{picture}
\figcaption[fig4.eps]{
Illustration of the effect of the different positron distributions in
the muon rest system.  The thick solid lines show the production spectrum
of secondary positrons and electrons per hydrogen atom for the
cosmic-ray proton spectrum given by Mori (1997).  The positron spectrum
with no kaon contribution is shown by the dash-dotted line.  The dashed
line shows the spectrum calculated assuming an isotropic distribution of
$e^+$ in the muon rest system ($\xi=0$), the dotted line shows the
spectrum calculated for $\xi=-1$.  The thin solid line with hatched regions
shows the positron production rate by cosmic rays in interstellar medium
(including contribution of He nuclei) from Protheroe (1982).
\label{fig4}}
\end{figure} 

\begin{figure}[thb]
   \begin{picture}(165,80)(0,0)
      \put(0,0){\makebox(85,0)[lb]%
{\psfig{file=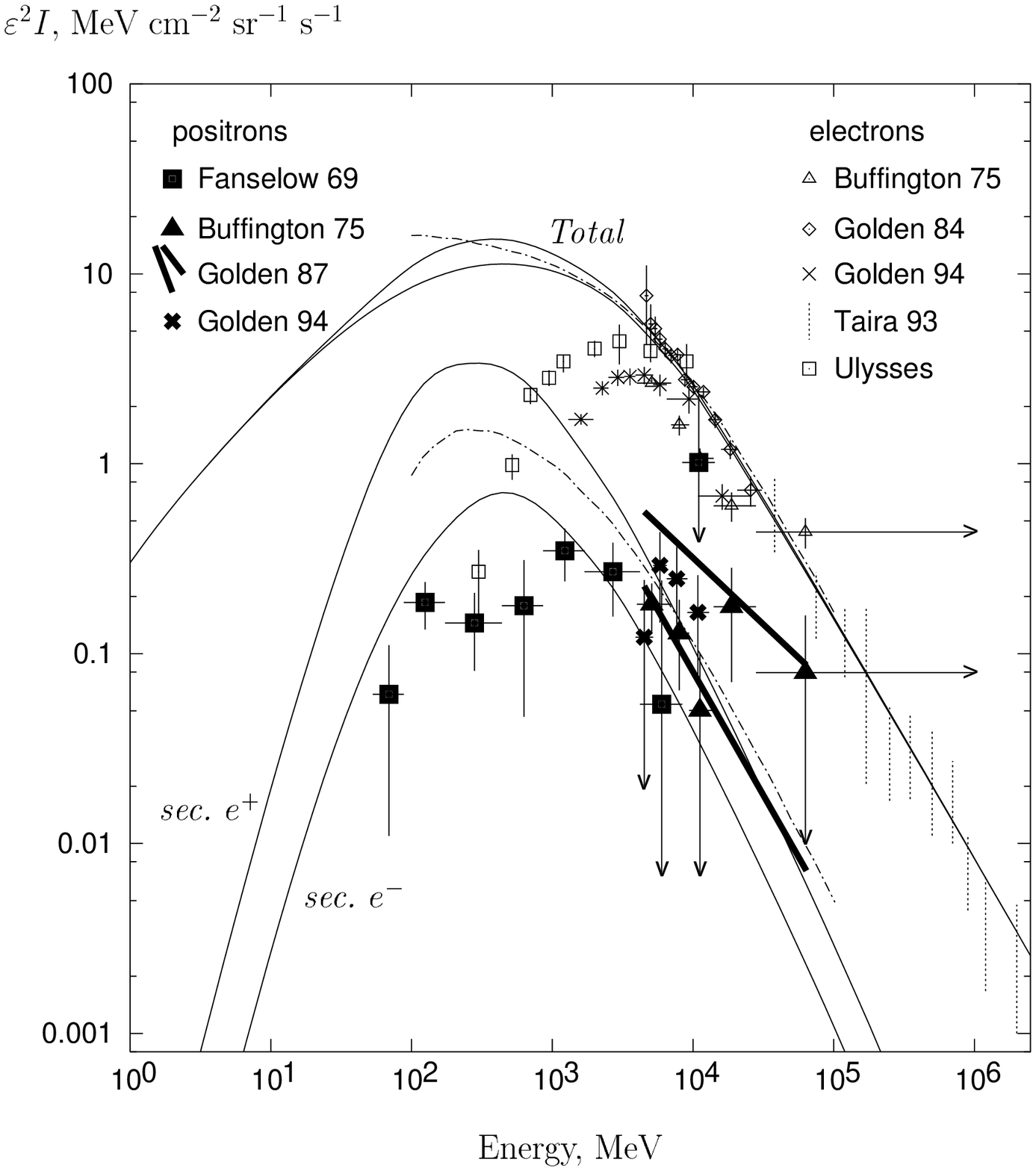,width=\figwidth,clip=}}}
      \put(85,0){\makebox(85,0)[lb]%
{\psfig{file=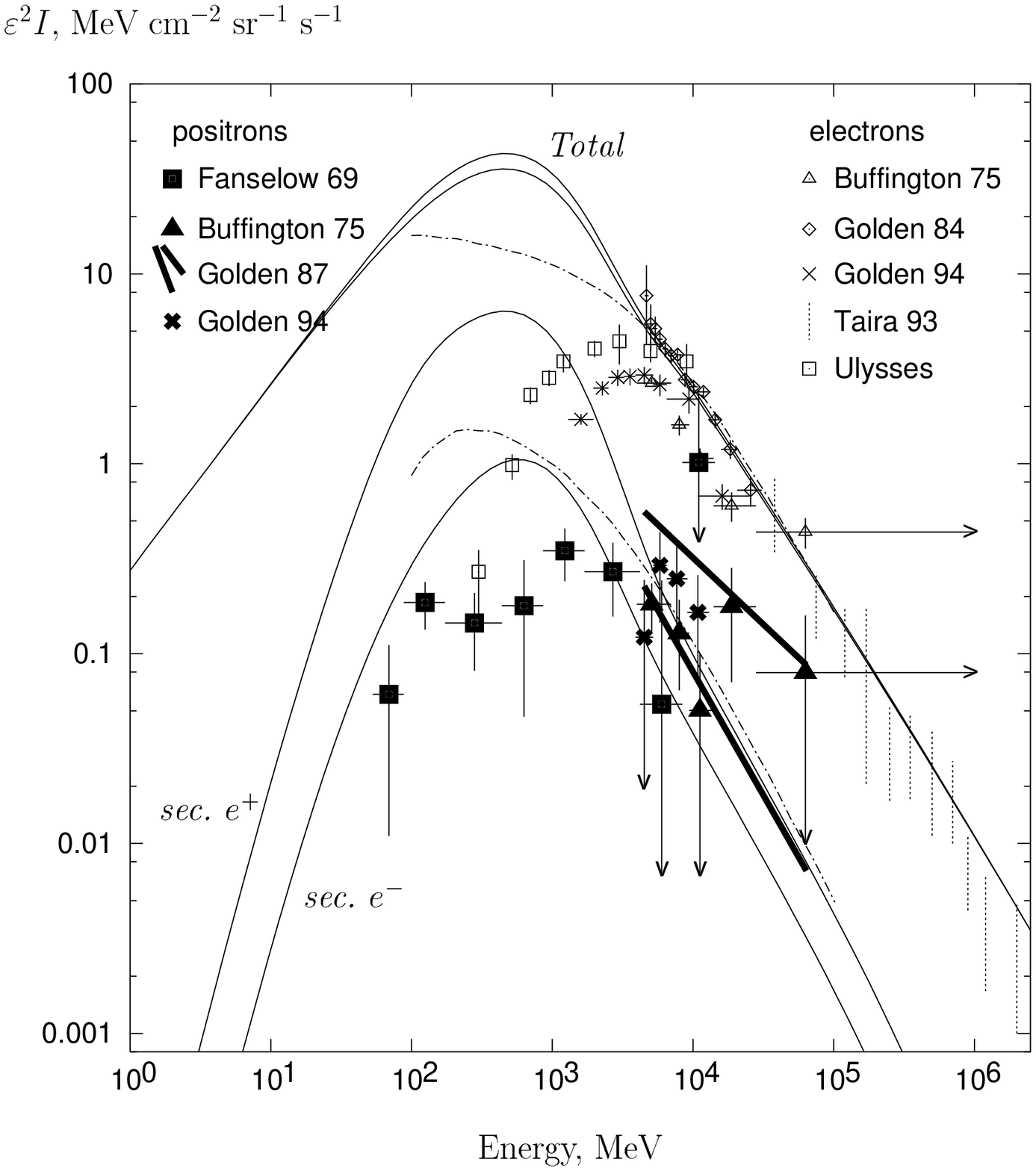,width=\figwidth,clip=}}}
   \end{picture}
\figcaption[fig5a.eps,fig5b.eps]{
Spectra of secondary positrons and electrons, and of primary electrons.
{\it Left panel}: model with no reacceleration (08--005). Electron
injection spectrum index 2.1, 2.4 above and below 10 GeV respectively.
Model:  upper curves: primary electrons and primary+secondary electrons
and positrons. Lower curves: secondary positrons, electrons. Lower
dashed-dot line: Protheroe (1982) leaky-box prediction.  Data:
electrons: Buffington, Orth, \& Smoot (1975), Golden et al.\ (1984),
Golden et al.\ (1994), Taira et al.\ (1993), Ulysses
(\cite{Ferrando96}), upper dashed-dot line: Protheroe (1982);
positrons:  Fanselow et al.\ (1969), Buffington, Orth, \& Smoot (1975),
Golden et al.\ (1987), Golden et al.\ (1994).  {\it Right panel}: same,
model with reacceleration (08--006).
\label{fig5}}
\end{figure} 

\begin{figure}[thb]
   \begin{picture}(165,80)(0,0)
      \put(0,0){\makebox(85,0)[lb]%
{\psfig{file=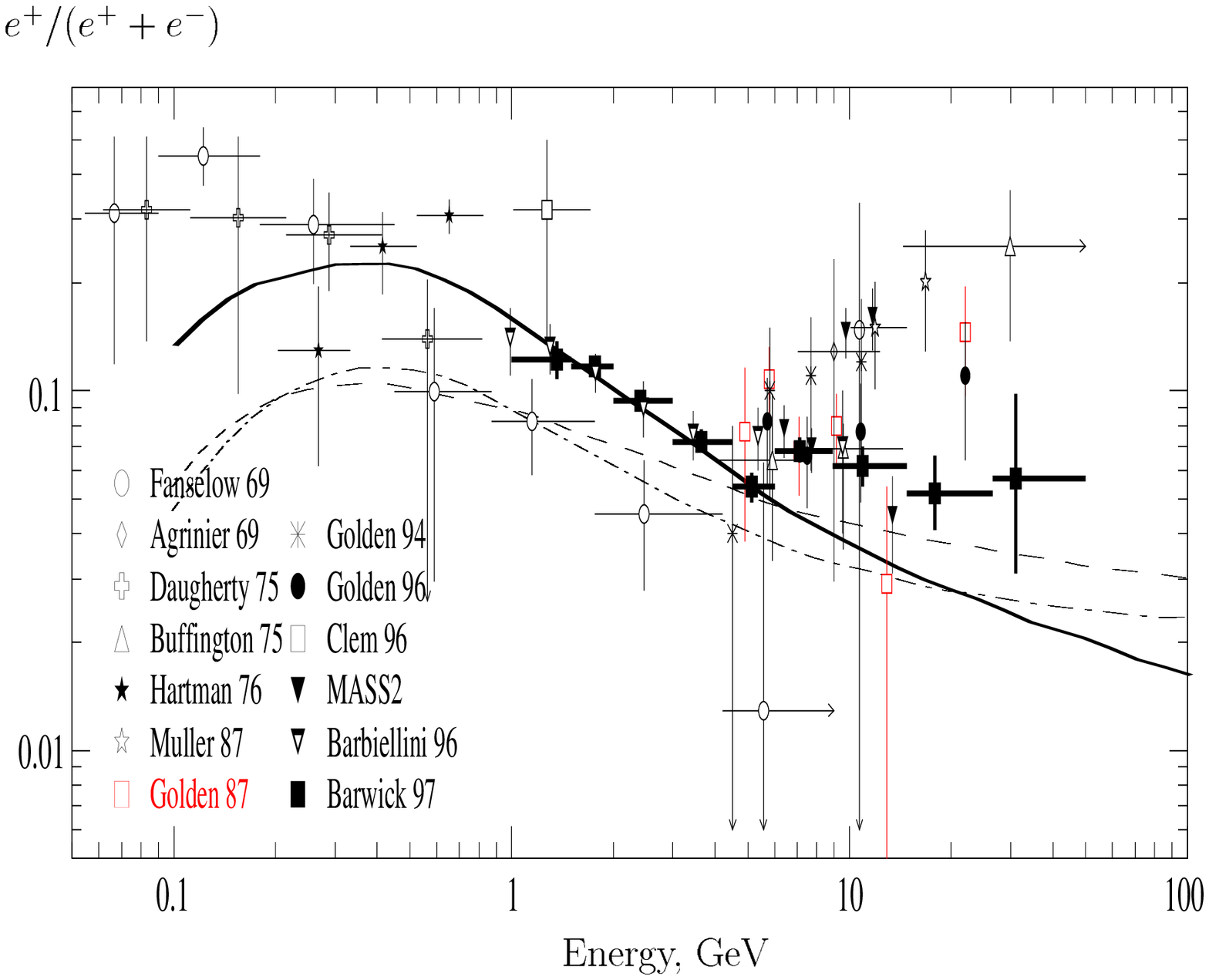,width=\figwidth,clip=}}}
      \put(85,0){\makebox(85,0)[lb]%
{\psfig{file=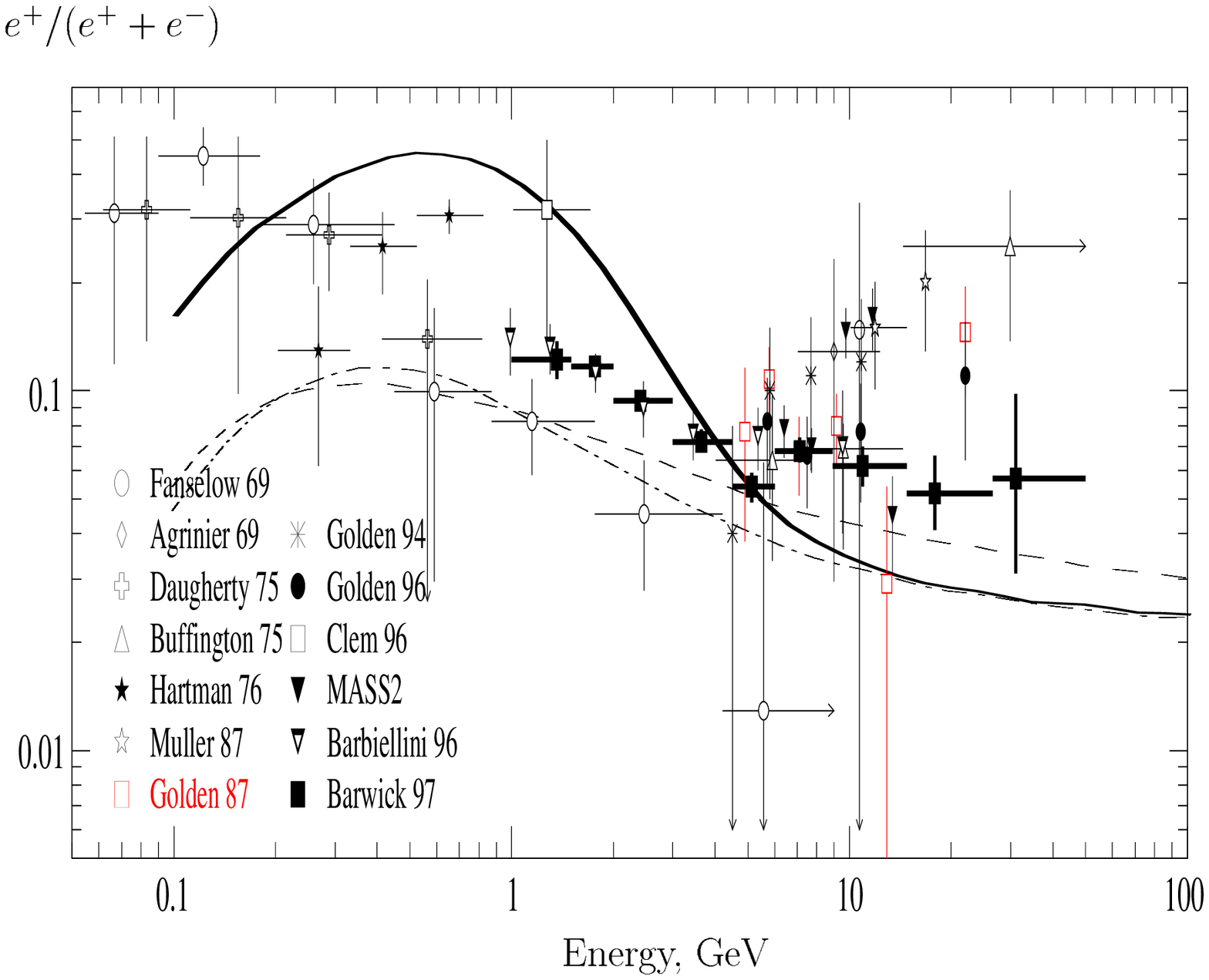,width=\figwidth,clip=}}}
   \end{picture}
\figcaption[fig6a.eps,fig6b.eps]{
{\it Left panel}: Positron fraction for model with no reacceleration. The
positron spectrum is divided by the electron spectrum used by Protheroe
(1982).  Dot-dashed line: positron fraction from Protheroe (1982),
leaky-box (dashed-dot), diffusion (dashed). The collection of the
experimental data is taken from Barwick et al.\ (1997).
{\it Right panel}: Same positron fraction for model with reacceleration.
\label{fig6}}
\end{figure} 

\begin{figure}[thb]
   \begin{picture}(165,80)(0,0)
      \put(0,0){\makebox(85,0)[lb]%
{\psfig{file=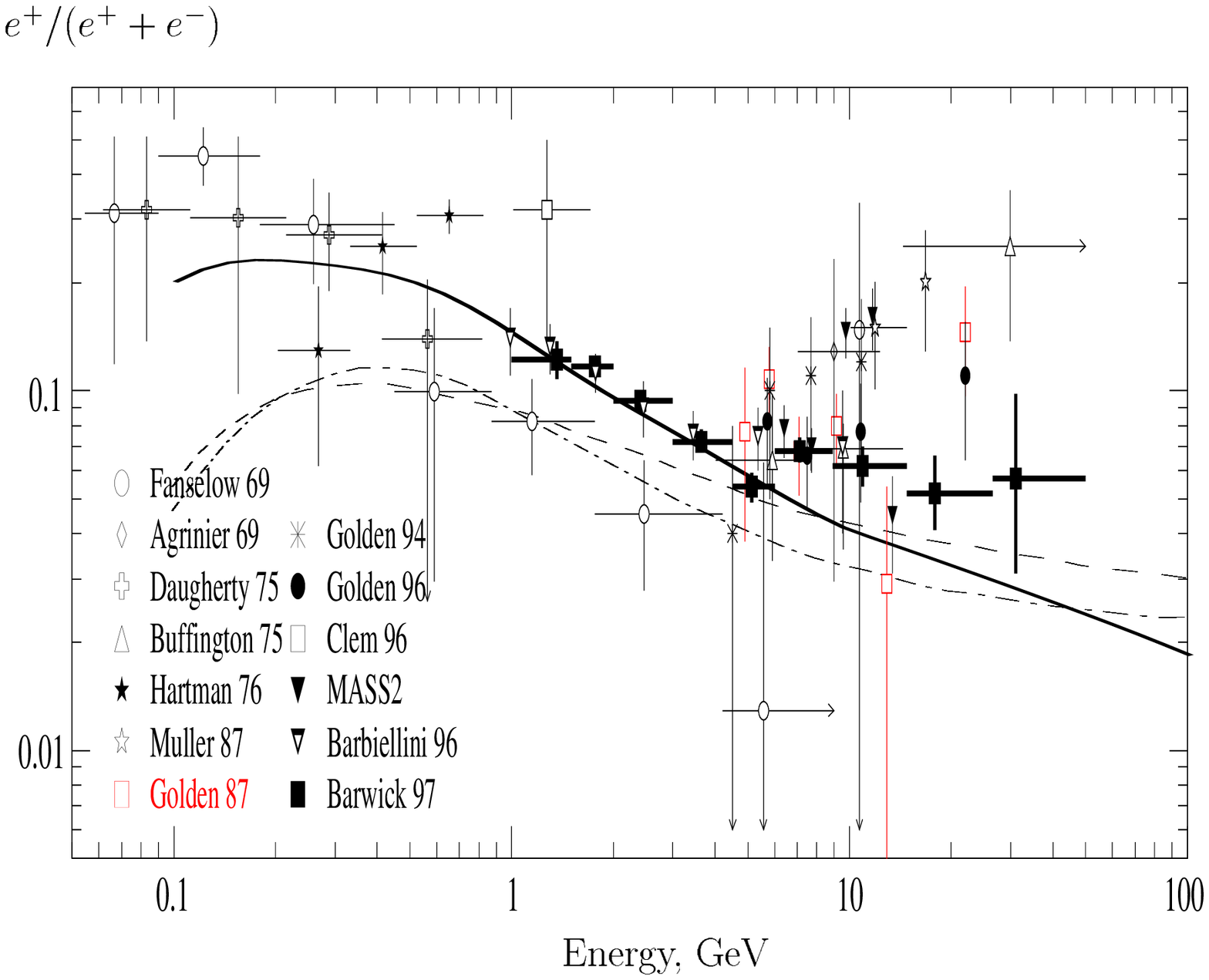,width=\figwidth,clip=}}}
      \put(85,0){\makebox(85,0)[lb]%
{\psfig{file=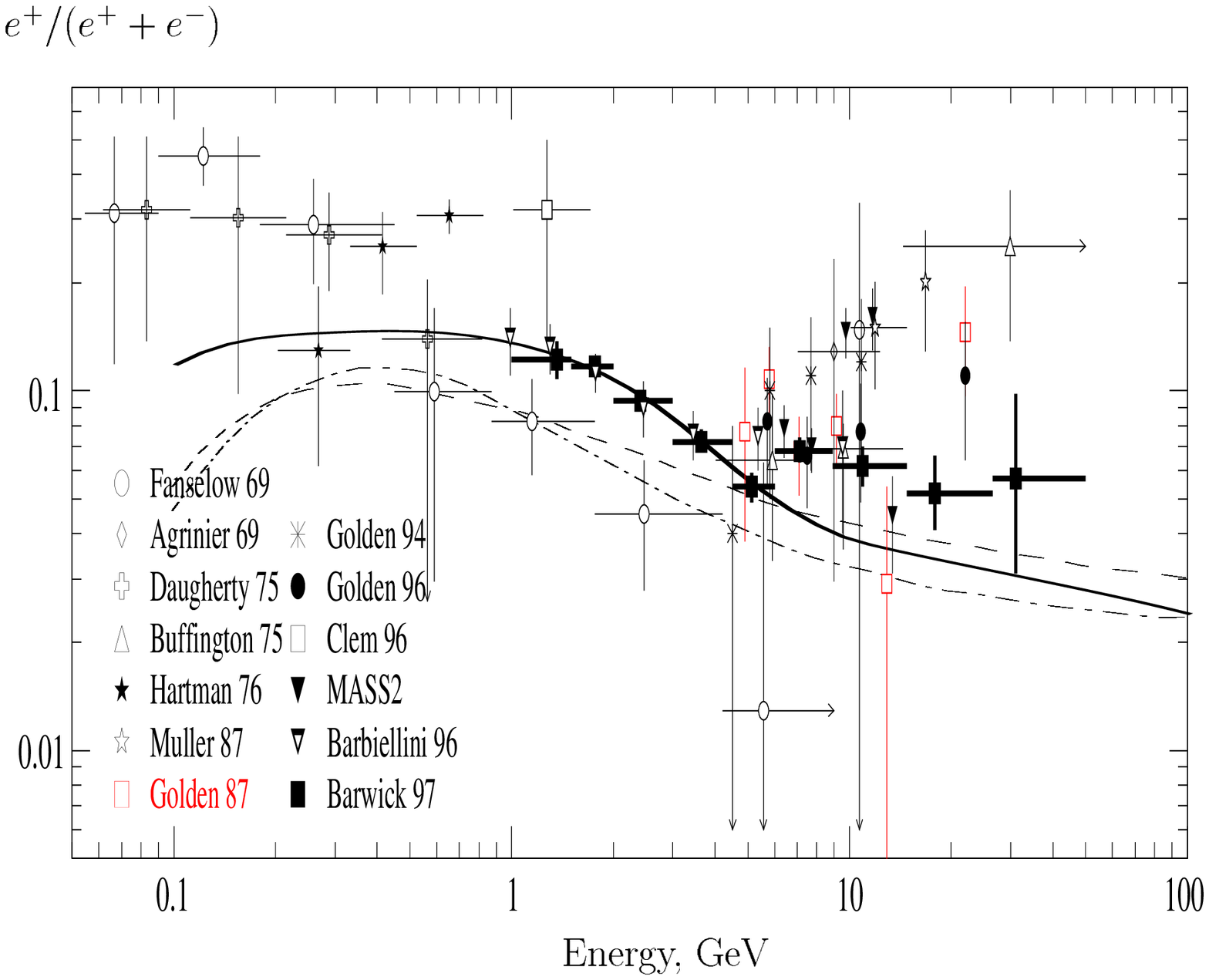,width=\figwidth,clip=}}}
   \end{picture}
\figcaption[fig7a.eps,fig7b.eps]{
{\it Left panel}: Positron fraction for  model with no reacceleration. The
positron spectrum is divided by the electron spectrum computed in the
propagation model. Data and other curves as Fig.~\ref{fig6}. {\it Right
panel}: Same positron fraction for model with reacceleration.  
\label{fig7}}
\end{figure} 

\begin{figure}[thb]
   \begin{picture}(165,80)(0,0)
      \put(0,0){\makebox(85,0)[lb]%
{\psfig{file=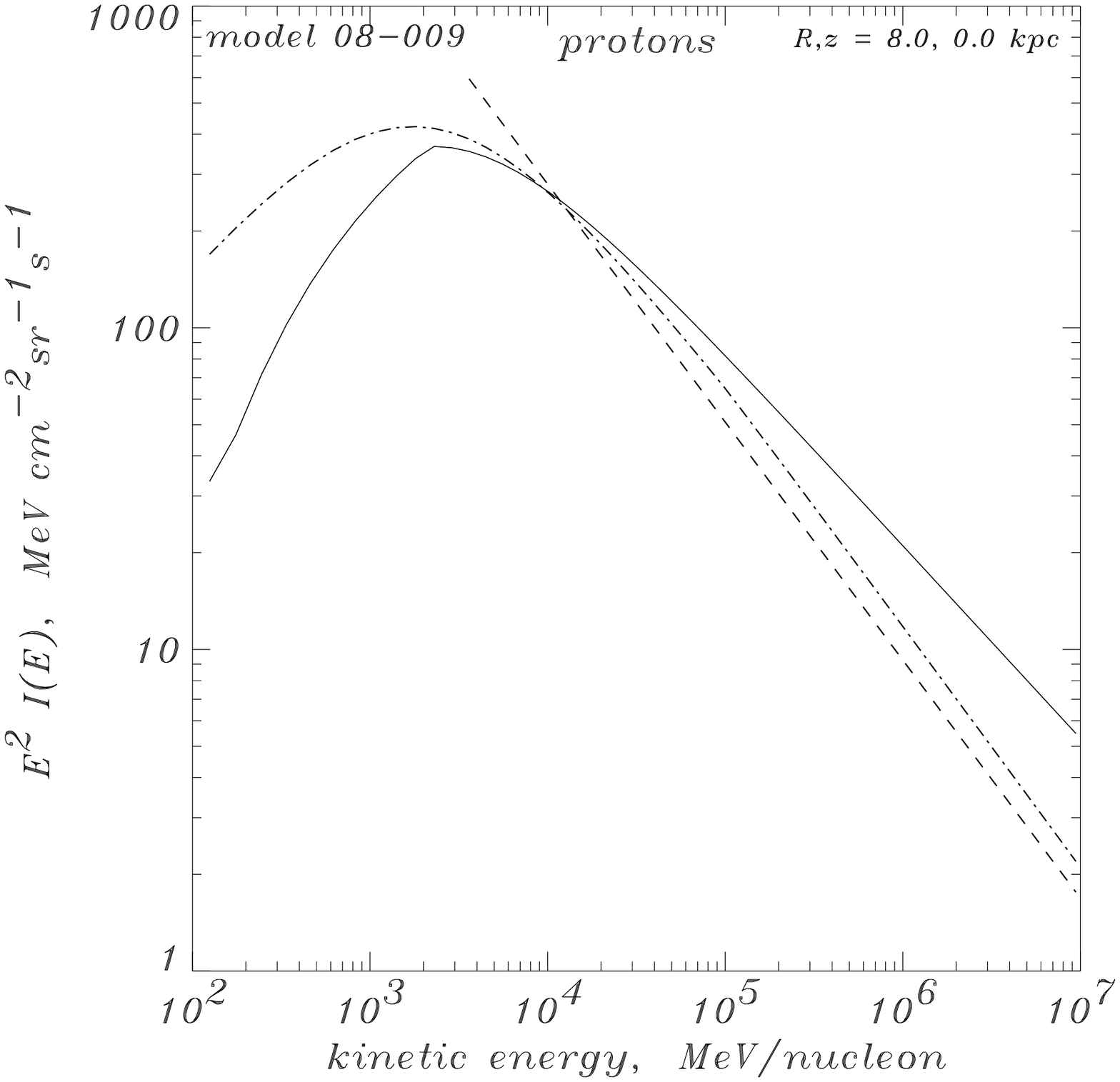,width=\figwidth,clip=}}}
      \put(85,0){\makebox(85,0)[lb]%
{\psfig{file=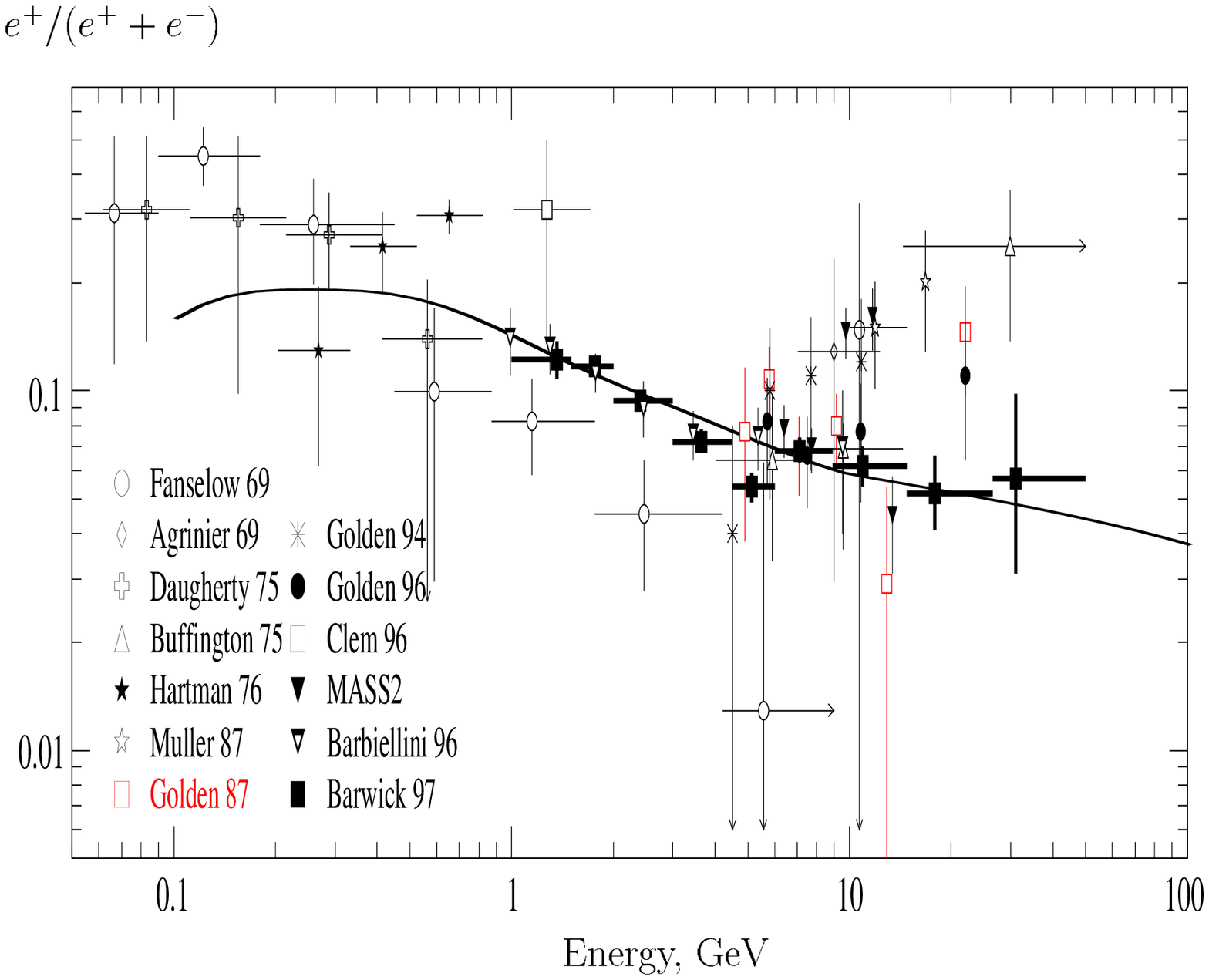,width=\figwidth,clip=}}}
   \end{picture}
\figcaption[fig8a.eps,fig8b.eps]{ 
{\it Left panel}:  spectrum of protons for a `flat' injection spectrum.
Thick line: model with no reacceleration (08--009), injection spectrum
index 2.0. Other spectra as in Fig.~\ref{fig2}.   {\it Right panel}:
positron fraction for this nucleon spectrum. Data as Fig.~\ref{fig6}.
\label{fig8}}
\end{figure} 


\clearpage

\begin{thebibliography}{}

\bibitem[Badhwar, Stephens, \& Golden 1977]{Badhwar77}
   Badhwar, G. D., Stephens, S. A., \& Golden, R. L.  1977, \prd, 15, 820

\bibitem[Barbiellini et al.\ 1996]{Barbiellini96}
   Barbiellini, G., et al.\  1996, \aap, 309, L15

\bibitem[Barwick et al.\ 1997]{Barwick97}
   Barwick, S. W., et al.\  1997, \apjl, 482, L191

\bibitem[Berezinskii et al.\ 1990]{Berezinskii90}
   Berezinskii, V. S., et al.\  1990, Astrophysics of Cosmic Rays
   (Amsterdam: North Holland)

\bibitem[Bloemen, Deul, \& Thaddeus 1990]{BloemenDeulThaddeus90}
   Bloemen, H., Deul, E. R., \& Thaddeus, P.  1990, \aap, 233, 437

\bibitem[Boulanger \& D\'esert 1992]{Boulanger92}
  Boulanger, F., \& D\'esert F. X.  1992, `The Infrared and
  Submillimetre Sky after COBE', M. Signore and C. Dupraz (eds), Kluwer
  Academic Publishers, Dordrecht, p. 263

\bibitem[Bronfmann et al.\ 1988]{Bronfmann88}
   Bronfmann, L., et al.\  1988, \apj, 324, 248

\bibitem[Buffington, Orth, \& Smoot 1975]{Buffington75}
   Buffington, A., Orth, C. D., \& Smoot, G. F.  1975, \apj, 199, 669

\bibitem[Clem et al.\ 1996]{Clem96}
   Clem, J. M., et al.\  1996, \apj, 464, 507

\bibitem[Cox, Kr\"ugel, \& Mezger 1986]{Cox86}
  Cox, P., Kr\"ugel, E., \& Mezger, P. G.  1986, \aap, 155, 380

\bibitem[Dermer 1986a]{Dermer86a}
   Dermer, C. D.  1986a, \apj, 307, 47

\bibitem[Dermer 1986b]{Dermer86b}
   Dermer, C. D.  1986b, \aap, 157, 223

\bibitem[Engelmann et al.\ 1985]{Engelmann85}
   Engelmann, J. J., et al.\  1985, \aap, 148, 12

\bibitem[Engelmann et al.\ 1990]{Engelmann90}
   Engelmann, J. J., et al.\  1990, \aap, 233, 96

\bibitem[Fanselow et al.\ 1969]{Fanselow69}
   Fanselow,  J. L., et al.\  1969, \apj, 158, 771

\bibitem[Ferrando et al.\ 1996]{Ferrando96}
   Ferrando, P., et al.\  1996, \aap, 316, 528

\bibitem[Gleeson \& Axford 1968]{GleesonAxford68}
   Gleeson, L. J., \& Axford, W. I.  1968, \apj, 154, 1011

\bibitem[Golden et al.\ 1984]{Golden84}
   Golden, R. L., et al.\  1984, \apj, 287, 622

\bibitem[Golden et al.\ 1987]{Golden87}
   Golden, R. L., et al.\  1987, \aap, 188, 145

\bibitem[Golden et al.\ 1994]{Golden94}
   Golden, R. L., et al.\  1994, \apj, 436, 769

\bibitem[Golden et al.\ 1996]{Golden96}
   Golden, R. L., et al.\  1996, \apjl, 457, L103 

\bibitem[Gordon \& Burton 1976]{GordonBurton76}
  Gordon, M. A., \& Burton, W. B.  1976, \apj, 208, 346

\bibitem[Gralewicz et al.\ 1997]{Gralewicz97}
   Gralewicz, P., et al.\  1997, \aap, 318, 925

 \bibitem[Heiles 1996]{Heiles96}
  Heiles, C.  1996, \apj, 462, 316 

\bibitem[Heinbach \& Simon 1995]{HeinbachSimon95}
   Heinbach, U., \& Simon, M.  1995, \apj, 441, 209

\bibitem[Hunter et al.\ 1997]{Hunter97}
   Hunter, S. D., et al.\  1997, \apj, 481, 205

\bibitem[Letaw, Silberberg, \& Tsao 1993]{Letaw93}
   Letaw, J. R., Silberberg, R., \& Tsao, C. H.  1993, \apj, 414, 601

\bibitem[Lukasiak et al.\ 1994]{Lukasiak94}
   Lukasiak, A., Ferrando, P., McDonald F. B., Webber, W. R.  1994,
   \apj, 423, 426

\bibitem[Mannheim \& Schlickeiser 1994]{MannheimSchlickeiser94}
   Mannheim, K., \& Schlickeiser, R.  1994, \aap, 286, 983

\bibitem[Mori 1997]{Mori97}
   Mori, M.  1997, \apj, 478, 225

\bibitem[Murphy, Dermer, \& Ramaty 1987]{Murphy87}
   Murphy, R. J., Dermer, C. D., \& Ramaty, R.  1987, \apjs, 63, 721

\bibitem[Orth \& Buffington 1976]{OrthBuffington76}
   Orth, C. D., \& Buffington, A.  1976, \apj, 206, 312

\bibitem[Particle Data Group 1990]{ParticleData90}
   Particle Data Group  1990, Phys. Lett. B, 239 

\bibitem[Phillipps et al.\  1981]{Phillipps81}
   Phillipps, S., et al. 1981, \aap, 103, 405

\bibitem[Porter \& Protheroe 1997]{PorterProtheroe97}
   Porter, T. A., \& Protheroe R. J. 1997, J. Phys. G., in press
   (astro-ph/9608182)

\bibitem[Protheroe 1982]{Protheroe82}
   Protheroe, R. J.  1982, \apj, 254, 391

\bibitem[Seo et al.\ 1991]{Seo91}
   Seo, E. S., et al.\  1991, \apj, 378, 763

\bibitem[Seo \& Ptuskin 1994]{SeoPtuskin94}
   Seo, E. S., \& Ptuskin, V. S.  1994, \apj, 431, 705

\bibitem[Simon \& Heinbach 1996]{SimonHeinbach96}
   Simon, M., \& Heinbach, U.  1996, \apj, 456, 519 

\bibitem[Stecker 1970]{Stecker70}
   Stecker, F. W.  1970, \apss, 6, 377

\bibitem[Stephens \& Badhwar 1981]{StephensBadhwar81}
   Stephens, S. A., \& Badhwar, G. D.  1981, \apss, 76, 213

\bibitem[Strong 1996]{Strong96}
   Strong, A. W.  1996, Spa. Sci. Rev., 76, 205

\bibitem[Strong et al.\ 1996]{Strongetal96}
   Strong, A. W., et al.\  1996, \aap{S}, 120, C381

\bibitem[Strong \& Mattox 1996]{StrongMattox96}
   Strong, A. W., \& Mattox, J. R.  1996, \aap, 308, L21

\bibitem[Strong \& Moskalenko 1997a]{StrongMoskalenko97a}
   Strong, A. W., \& Moskalenko, I. V.  1997a, 4th Compton Symp., AIP,
   in press (astro-ph/9709211)

\bibitem[Strong \& Moskalenko 1997b]{StrongMoskalenko97b}
   Strong, A. W., \& Moskalenko, I. V.  1997b, in preparation

\bibitem[Strong, Moskalenko, \& Sch\"onfelder 1997]{Strong97}
   Strong, A. W., Moskalenko, I. V., \& Sch\"onfelder, V.  1997, 
   25th ICRC (Durban), 4, 257 (astro-ph/9706010)

\bibitem[Strong \& Youssefi 1995]{StrongYoussefi95}
   Strong, A. W., \& Youssefi, G.  1995, Proc. 24th ICRC (Roma), 3, 48

\bibitem[Taira et al.\ 1993]{Taira93}
   Taira, T., et al.\  1993, Proc. 23rd ICRC (Calgary), 2, 128

\bibitem[Vall\'ee  1994]{Vallee94}
   Vall\'ee, J. P. 1994, \apj, 437, 179

\bibitem[Webber et al.\ 1996]{Webber96}
   Webber, W. R., et al.\  1996, \apj, 457, 435

\bibitem[Webber, Lee, \& Gupta 1992]{Webber92}
   Webber, W. R., Lee, M. A., \& Gupta, M.  1992, \apj, 390, 96

\end{thebibliography}
\end{document}